\documentclass[11pt]{article}
\pdfoutput=1

\usepackage{jheppub}
\usepackage{amssymb,amsfonts,amsmath,amsthm,lineno}
\usepackage{enumerate}
\usepackage{mathrsfs}
\usepackage{bbold}
\usepackage{tikz}
\usetikzlibrary{shapes.geometric, arrows}
\usepackage{hyperref}
\usepackage{slashed}
\usepackage{color}
\usepackage{empheq}

\makeatletter
\newcommand{\Vast}{\bBigg@{4.75}}
\makeatother

\newcommand{\be}{\begin{equation}}
\newcommand{\ee}{\end{equation}}
\newcommand{\bea}{\begin{eqnarray}}
\newcommand{\eea}{\end{eqnarray}}

\newcommand{\CA}{\mathcal{A}}

\newcommand{\CC}{\mathcal{C}}

\newcommand{\CF}{\mathcal{F}}

\newcommand{\CK}{\mathcal{K}}
\newcommand{\CL}{\mathcal{L}}

\newcommand{\CM}{\mathcal{M}}

\newcommand{\lr}{\left (}
\newcommand{\rr}{\right )}
\newcommand{\ls}{\left [}
\newcommand{\rs}{\right ]}

\newcommand\qt\tau

\newcommand{\Lim}[1]{\raisebox{0.5ex}{\scalebox{1}{$\displaystyle \lim_{#1}\;$}}}

\newcommand{\p}{\partial}

\renewcommand{\tilde}[1]{\widetilde{#1}}
\newcommand{\tr}{\text{tr}}

\makeatletter
\renewcommand{\@seccntformat}[1]{\csname the#1\endcsname.\,\,}
\makeatother
\allowdisplaybreaks

\DeclareMathOperator{\sgn}{sgn}

\let \savenumberline \numberline
\def \numberline#1{\savenumberline{#1.}}

\makeatletter
\def\@fpheader{\relax}
\makeatother

\def\bea{\begin{eqnarray}}
\def\eea{\end{eqnarray}}

\usetikzlibrary{shapes,arrows,chains}
\usetikzlibrary{decorations.markings}
\usetikzlibrary{decorations.pathmorphing}
\usetikzlibrary{shapes.multipart}
\tikzset{snake it/.style={decorate, decoration=snake}}

\usepackage{tikz-cd}
\usetikzlibrary{cd,decorations.markings, arrows.meta}

\newcommand{\SLR}{SL(2,\,$\mathbb{R}$)}
\newcommand{\SLZ}{SL(2,\,$\mathbb{Z}$)}

\newcommand{\Cz}{C^{(0)}}

\newcommand{\hS}{\hat{\Sigma}}

\title{\ \vspace{1.6cm} \\
Branched \texorpdfstring{SL(2,\,$\mathbb{Z}$)}{Z} Duality}
\author[a]{Eric A. Bergshoeff,}
\author[b]{Kevin T. Grosvenor,}
\author[a]{Johannes Lahnsteiner,}
\author[c]{Ziqi Yan,}
\author[d]{\\and Utku Zorba}
\emailAdd{e.a.bergshoeff@rug.nl}
\emailAdd{kevinqg1@gmail.com}
\emailAdd{j.m.lahnsteiner@outlook.com}
\emailAdd{ziqi.yan@su.se}
\emailAdd{utku.zorba@boun.edu.tr}
\affiliation[a]{Van Swinderen Institute, University of Groningen\\
Nijenborgh 4, 9747 AG Groningen, The Netherlands \smallskip
}
\affiliation[b]{Instituut-Lorentz, Universiteit Leiden \\
P.O. Box 9506, 2300 RA Leiden, The Netherlands \smallskip
}
\affiliation[c]{
Nordita, KTH Royal Institute of Technology and Stockholm University\\
Hannes Alfv\'{e}ns v\"{a}g 12, SE-106 91 Stockholm, Sweden \smallskip}
\affiliation[d]{
Physics Department, Bo\u{g}azi\c{c}i University\\
34342 Bebek, Istanbul, Turkey
}
\abstract{We investigate how SL(2,\,$\mathbb{Z}$) duality is realized in nonrelativistic type IIB superstring theory, which is a self-contained corner of relativistic string theory. Within this corner, we realize manifestly \SLZ-invariant $(p\,,q)$-string actions. The construction of these actions imposes a branching between strings of opposite charges associated with the two-form fields. The branch point is determined by these charges and the axion background field. Both branches must be incorporated in order to realize the full SL(2,\,$\mathbb{Z}$) group.
Besides these string actions, we also construct D-instanton and D3-brane actions that manifestly realize the branched SL(2,\,$\mathbb{Z}$) symmetry.
}

\begin{document}

\maketitle
\vfill\eject

\section{Introduction}

Dualities have played a key role in probing the different non-perturbative corners of string/M-theory. In particular, type IIB superstring theory exhibits an SL(2,\,$\mathbb{Z}$) symmetry connecting its perturbative and non-perturbative sectors. This SL(2,\,$\mathbb{Z}$) symmetry survives in the low-energy supergravity limit, which greatly facilitates the study of the spectrum of supergravity solutions.\footnote{Historically, the discovery went in the other direction: SU(1,1)\,$\cong$\,\SLR\ was observed as a global symmetry of type IIB supergravity first
in \cite{Schwarz:1983wa}. A decade later, this symmetry group
was finally proposed as a symmetry in type IIB superstring theory in \cite{Hull:1994ys} (also see \cite{Schwarz:1995dk}).
See \emph{e.g.} page 695 of \cite{Becker:2006dvp} for a review of the history. See also \cite{cardy1982duality, cardy1982phase, shapere1989self} for earlier references on the SL(2,\,$\mathbb{Z}$) duality in lattice models. We thank Paul Townsend for pointing out this early occurrence of the SL(2,\,$\mathbb{Z}$) duality.} A realization of this SL(2,\,$\mathbb{Z}$) symmetry was given in the context of the string sigma model by \cite{Townsend:1997kr, Cederwall:1997ts} and in the context of the effective worldvolume actions of D-branes by \cite{Tseytlin:1996it,Bergshoeff:2006gs}.
D-branes are useful probes for understanding how the SL(2,\,$\mathbb{Z}$) group acts on the background fields of type IIB superstring theory. In this paper, we will follow this approach as our guiding principle and use D-branes to study the SL(2,\,$\mathbb{Z}$) duality in the less well-understood case of nonrelativistic string theory \cite{Klebanov:2000pp, Gomis:2000bd, Danielsson:2000gi}.

Recently, a systematic construction of the non-perturbative duality web in nonrelativistic string/M-theory has been initiated in \cite{Ebert:2021mfu}. Nonrelativistic string theory arises as a corner of relativistic string theory, under a zero Regge slope limit where the Kalb-Ramond field is fine-tuned to its critical value that matches the string tension \cite{Klebanov:2000pp, Gomis:2000bd, Danielsson:2000gi}. This leads to a string theory with a spectrum satisfying a Galilean-invariant dispersion relation. This corner defines a self-consistent and ultra-violet complete string theory on its own \cite{Gomis:2000bd}. Furthermore, it has been established that this theory provides a first-principles definition for string theory in the discrete light cone quantization (DLCQ) \cite{Bergshoeff:2018yvt}. DLCQ plays an important role in nonperturbative approaches to string/M-theory such as Matrix theory \cite{Banks:1996vh, Susskind:1997cw, Seiberg:1997ad, Sen:1997we}.

Lately, effective field theories that arise from nonrelativistic string theory have been intensively studied, leading to novel nonrelativistic gravity and gauge theories \cite{Andringa:2012uz, Harmark:2017rpg, Harmark:2018cdl, Harmark:2019upf, Bergshoeff:2019pij, Kluson:2019ifd, Kluson:2019avy, Gallegos:2020egk, Gomis:2020fui, Bergshoeff:2021bmc, Bidussi:2021ujm, Bergshoeff:2021tfn, Bergshoeff:2022pzk} (for further references, see the review \cite{Oling:2022fft}). In \cite{Andringa:2012uz, Bergshoeff:2018yvt}, it is shown that the spacetime geometry underlying nonrelativistic closed string theory is non-Riemannian and has a codimension-two foliation structure.\,\footnote{Here, we have a slight abuse of the terminology ``foliation," which is technically only a ``distribution" in mathematics. This distinction is not important in this paper.} The spacetime equations of motion that govern the dynamics of the target-space geometry have been studied in \cite{Gomis:2019zyu, Gallegos:2019icg, Bergshoeff:2019pij, Yan:2019xsf, Yan:2021lbe}
by imposing quantum Weyl invariance. Extending the discussion to open strings, the effective worldvolume actions for D-branes coupled to the bosonic closed string background fields including the Ramond-Ramond fields have been studied in \cite{Gomis:2020fui, Ebert:2021mfu}. This provides a framework that unifies nonrelativistic Yang-Mills theory \cite{Gomis:2020fui}, noncommutative open string theory and noncommutative Yang-Mills theory \cite{Gopakumar:2000na}. Analyses of various T- and S-duality transformations of nonrelativistic D-branes have appeared in \cite{Kluson:2019avy, Gomis:2020izd, Ebert:2021mfu}. In particular, a nonrelativistic $(p\,, q)$-string action has been derived in \cite{Ebert:2021mfu}.\,\footnote{Here, $p$ and $q$ are two coprime integers that transform as a doublet under the SL(2,\,$\mathbb{Z}$) duality symmetry. We use the convention where $(p\,,q) = (1,0)$ corresponds to the fundamental string.}

So far, the full SL(2,\,$\mathbb{Z}$) symmetry in nonrelativistic string theory has not yet been thoroughly studied.\,\footnote{See \emph{e.g.} \cite{Russo:2000zb, Cai:2000yk, Lu:2000vv, Gran:2001tk} for studies of SL(2,\,$\mathbb{Z}$) duality in the context of noncommutative open string theory.} It is the purpose of this paper to fill this gap and to investigate how this SL(2,\,$\mathbb{Z}$) duality manifests itself in the context of nonrelativistic string theory. Surprisingly, the SL(2,\,$\mathbb{Z}$) symmetry is realized in a nontrivial way where a branching arises, which splits the $(p\,,q)$ parameter space into two halves. The SL(2,\,$\mathbb{Z}$) transformations connect these two branches. Crucially, the branching is characterized not only by the values of $p$ and $q$, but also by the position $x$ in spacetime via the background axion Ramond-Ramond field $\Cz (x)$, through the sign of the quantity
\begin{equation}
    p - q \, \Cz\,.
\end{equation}
Note that this condition itself transforms under SL(2,\,$\mathbb{Z}$) , which therefore splits this global symmetry group into two halves, in an $x$-dependent way, to those that preserve and those that flip the sign of $p - q \, \Cz$. The inter-branch case where $p - q \, \Cz = 0$ requires careful attention and will be discussed separately in this paper.

The paper is organized as follows. In Section \ref{sec:nrstdb}, we review some essential ingredients in nonrelativistic string theory.
In Section~\ref{sec:nrpq}, we derive the nonrelativistic $(p\,, q)$-string action in Eqs.~\eqref{eq:spq} and \eqref{eq:spqm} and show that its $(p\,,q)$-space splits into two branches. We then derive the associated SL(2,\,$\mathbb{Z}$) transformations in Eq.~\eqref{eq:slztfs} (and Eq.~\eqref{eq:c4trnsf}), which are also branched.
At the end of this section, we study a tensionless limit of nonrelativistic $(p\,, q)$-strings with the inter-branch condition $p-q\,\Cz =0$\,.
In Section~\ref{sec:d3}, we extend our discussion to other nonrelativistic D$p$-branes in type IIB superstring theory, including the D-instanton and D3-brane described by the action \eqref{eq:sd3bb}. We also study the inter-branch case for the nonrelativistic D3-brane action with $p - q \, \Cz = 0$\,. We conclude the paper in Section~\ref{sec:concl}. This paper has two appendices. In Appendix~\ref{app:ivs}, we present a systematic derivation of the \SLZ-invariants in nonrelativistic string theory. In Appendix~\ref{sec:NRlimits}, we give an alternative derivation of the main results of this paper via a nonrelativistic limit of relativistic string theory.

\section{Nonrelativistic String Theory and D-Branes}\label{sec:nrstdb}

We start with a review of some basic ingredients of nonrelativistic string theory and the relevant D$p$-brane actions, which will be essential for our later constructions of manifestly \SLZ-invariant D$p$-brane actions in nonrelativistic string theory.

\subsection{Nonrelativistic Closed String Theory} \label{sec:reviewncs}

Nonrelativistic closed string theory is defined by a two-dimensional (relativistic) sigma model that maps the worldsheet $\Sigma$ to a $d$-dimensional spacetime manifold $\CM$\,. Define the worldsheet coordinates to be $\sigma^\alpha$, $\alpha = 0\,, 1$ and the worldsheet metric $h_{\alpha\beta}$\,. In addition to the worldsheet fields $X^\mu (\sigma)$\,, $\mu = 0\,, \cdots, d-1$ representing the spacetime coordinates, the sigma model describing nonrelativistic strings also contains a pair of one-form fields $\lambda$ and $\bar{\lambda}$. Moreover, it is useful to introduce the zweibein field $e_\alpha{}^a$\,, $a = 0,1$ associated with $h_{\alpha\beta}$\,, such that $h_{\alpha \beta} = e_\alpha{}^a \, e_\beta{}^b \, \eta_{ab}$\,. In flat spacetime, the sigma model is given by \cite{Gomis:2000bd, Bergshoeff:2018yvt}
\be \label{eq:saction}
    S = - \frac{1}{4\pi\alpha'} \int d^2 \sigma \ls \sqrt{-h} \, h^{\alpha\beta} \, \p_\alpha X^{A'} \, \p_\beta X^{B'} \, \delta_{A' B'} + \epsilon^{\alpha\beta} \bigl( \lambda \, e_\alpha \, \p_\beta X + \bar{\lambda} \, \bar{e}_\alpha \, \p_\beta \overline{X} \, \bigr) \rs,
\ee
where $\alpha'$ is the Regge slope, $e_\alpha = e_\alpha{}^0 + e_\alpha{}^1$ and $\bar{e}_\alpha = e_\alpha{}^0 - e_\alpha{}^1$. Moreover, $X = X^0 + X^1$ and $\overline{X} = X^0 - X^1$ are the lightlike coordinates in the longitudinal sector of the spacetime manifold $\CM$\,. The index $A' = 2, \cdots, d-1$ represents the transverse sector. Since the transverse sector is of Euclidean signature, we will not distinguish upper and lower transverse indices. Locally, in the operatorial formalism, $\lambda$ and $\bar{\lambda}$ are associated with derivatives of the T-dual coordinates of $X$ and $\overline{X}$\,, respectively \cite{Yan:2021lbe}. While $X$ and $\overline{X}$ are conjugate to the momenta, the T-dual coordinates are conjugate to the windings. The closed string sector has a nontrivial spectrum if the longitudinal spatial direction $X^1$ is compactified over a circle of radius $R$\,. The dispersion relation and the level matching condition are \cite{Gomis:2000bd}
\be
    E = \frac{\alpha'}{2 w R} \ls K^{A'} K^{A'} + \frac{2}{\alpha'} \bigl( N + \tilde{N} - 2 \bigr) \rs,
        \qquad%
    \tilde{N} - N = n \, w\,.
\ee
Here, $E$ is the energy, $K^{A'}$ is the transverse momentum, $w$ is the winding number in $X^1$, $n$ is the Kaluza-Klein momentum number in the $X^1$ circle, and $N$ and $\tilde{N}$ are the string excitation numbers. T-dualizing nonrelativistic string theory along $X^1$ defines the discrete light cone quantization (DLCQ) of relativistic string theory, where the longitudinal spacelike circle is mapped to a lightlike circle \cite{Bergshoeff:2018yvt}.

In curved spacetime, the action \eqref{eq:saction} is generalized to be \cite{Gomis:2005pg, Bergshoeff:2018yvt}
\begin{align} \label{eq:genaction}
\begin{split}
    S & = - \frac{1}{4\pi\alpha'} \int d^2 \sigma \ls \sqrt{-h} \, h^{\alpha\beta} \, \p_\alpha X^{\mu} \, \p_\beta X^{\nu} \, E_{\mu\nu} + \epsilon^{\alpha\beta} \lr \lambda \, e_\alpha \, \p_\beta X^\mu \, \tau_\mu + \bar{\lambda} \, \bar{e}_\alpha \, \p_\beta X^\mu \, \bar{\tau}_\mu \, \rr \rs \\[2pt]
    & \quad - \frac{1}{4\pi\alpha'} \int d^2 \sigma \, \epsilon^{\alpha\beta} \, \p_\alpha X^\mu \, \p_\beta X^\nu \, B_{\mu\nu} + \frac{1}{4\pi} \int d^2 \sigma \sqrt{-h} \, R \, \Phi\,.
\end{split}
\end{align}
Here, we require that the symmetric two-tensor $E_{\mu\nu}$ take a restricted form with
\be
    E_{\mu\nu} = E_\mu{}^{A'} \, E_\nu{}^{A'}\,,
\ee
where $E_\mu{}^{A'}$ plays the role of the transverse vielbein in spacetime.\,\footnote{An arbitrary symmetric two-tensor $E_{\mu\nu}$ can always be brought into the form of $E_{\mu\nu} = E_\mu{}^{A'} E_\nu{}^{A'}$ by redefining the one-form fields $\lambda$ and $\bar{\lambda}$ together with the Kalb-Ramond field $B_{\mu\nu}\,$ \cite{Oling:2022fft}.} We also introduced the longitudinal vielbein $\tau_\mu{}^A$\,, $A = 0, 1$ and defined
\be
    \tau_\mu = \tau_\mu{}^0 + \tau_\mu{}^1\,,
        \qquad%
    \bar{\tau}_\mu = \tau_\mu{}^0 - \tau_\mu{}^1\,.
\ee
Finally, the theory is also coupled to the spacetime dilaton field $\Phi$ via the worldsheet Einstein-Hilbert term. Such geometrical data encoded by $\tau_\mu{}^A$ and $E_\mu{}^{A'}$\,, together with other background fields including the Kalb-Ramond field $B_{\mu\nu}$ and dilaton field $\Phi$\,, defines the so-called string Newton-Cartan geometry \cite{Andringa:2012uz}, which is endowed with a codimension-two foliation structure. This geometry naturally generalizes Newton-Cartan geometry with a codimension-one foliation structure for particles.
There is no graviton in the string spectrum, and the only gravitational force is the instantaneous Newton-like interaction between winding strings \cite{Gomis:2000bd, Danielsson:2000mu}.

The one-form fields $\lambda$ and $\bar{\lambda}$ in Eq.~\eqref{eq:genaction} play the role of Lagrange multipliers that impose the constraint equations,
\begin{align}
    \epsilon^{\alpha\beta} \, e_\alpha \, \tau_\beta = \epsilon^{\alpha\beta} \, \bar{e}_\alpha \, \bar{\tau}_\beta = 0\,,
\end{align}
where we have introduced the pullback $\tau_\alpha{}^A = \p_\alpha X^\mu \, \tau_\mu{}^A$.\,\footnote{When we use form notation, we do not distinguish between the worldvolume and target-space. It should be clear from the context which one we mean. } Up to an undetermined scale factor, these constraints are solved by
\be
    e_\alpha \propto \tau_\alpha\,,
        \qquad%
    \bar{e}_\alpha \propto \bar{\tau}_\alpha\,.
\ee
In the absence of the Einstein-Hilbert term, we use these equations to eliminate the auxiliary worldsheet metric $h_{\alpha\beta}$ in Eq.~\eqref{eq:genaction} to find the following Nambu-Goto action \cite{Andringa:2012uz}:
\be \label{eq:ngfsa}
    S_\text{NG} = - \frac{1}{4\pi\alpha'} \int d^2 \sigma \lr \sqrt{-\tau} \, \tau^{\alpha\beta} \, E_{\alpha\beta} + \epsilon^{\alpha\beta} \, B_{\alpha\beta} \rr.
\ee
Here, we have defined the longitudinal metric $\tau_{\mu\nu}\equiv \tau_\mu{}^A \, \tau_\nu{}^B \, \eta_{AB}$ and its pullback to the worldsheet $\tau_{\alpha\beta} = \p_\alpha X^\mu \, \p_\beta X^\nu \, \tau_{\mu\nu}$.
The metric determinant and inverse metric are denoted as $\tau = \det \tau_{\alpha\beta}$ and $\tau^{\alpha\beta}$, respectively. We also introduced the pullbacks $E_{\alpha\beta} = \p_\alpha X^\mu \, \p_\beta X^\nu \, E_{\mu\nu}$ and $B_{\alpha\beta} = \p_\alpha X^\mu \, \p_\beta X^\nu \, B_{\mu\nu}$\,.

\subsection{Open Strings and Nonrelativistic D-Branes}

In the case where the worldsheet $\Sigma$ has a boundary $\p\Sigma$\,, it is also possible to introduce open string vertex operators whose coherent states give rise to the boundary action. To be concrete, we map the worldsheet to the upper half plane, with the boundary $\p\Sigma$ identified with the real axis at $\sigma = 0$\,. Open strings have to end on D-branes. For simplicity, we consider the case of a single D$p$-brane, which is a submanifold embedded in the spacetime manifold $\CM$\,, parametrized by
\be
    X^\mu \big|_{\p \Sigma} = f^\mu (Y^\alpha)\,,
        \qquad%
    \alpha = 0\,, \, \cdots, \, p\,,
\ee
where $Y^\alpha$ are coordinates on the D-brane and $f^\mu$ is the embedding function. It is illuminating to consider a D$(d-2)$-brane, which is transverse to the longitudinal spatial direction and extends in the remaining directions. The covariant boundary action is \cite{Gomis:2020fui}
\be
    S_\text{b} = \frac{1}{2\pi\alpha'} \oint_{\p\Sigma} d\tau \Bigl[ \tfrac{1}{2} \lr \lambda - \bar{\lambda} \rr \nu + \p_\tau Y^\alpha A^\text{\scalebox{0.8}{B}}_\alpha \Bigr]\,,
        \qquad%
    \alpha = 0\,, 2\,, \cdots, d-1\,,
\ee
where $\nu$ corresponds to a scalar mode that perturbs the shape of the D-brane and $A^\text{\scalebox{0.8}{B}}_\alpha$ is a $U(1)$ gauge field on the D-brane. This theory describes nonrelativistic open strings \cite{Danielsson:2000mu}. String amplitudes in nonrelativistic open string theory have been recently studied in \cite{Yan:2021hte}, where a new KLT relation between nonrelativistic closed and open string amplitudes is realized.

In flat spacetime, and with the longitudinal spatial direction being compactified over a circle of radius $R$\,, the open string states enjoy a Galilean-invariant dispersion relation \cite{Danielsson:2000mu},
\be
    E = \frac{\alpha'}{2 \, w \, R} \ls K^{A'} \, K_{A'} + \frac{1}{\alpha'} \lr N - 1 \rr \rs.
\ee
Requiring that the theory is consistent at the quantum level sets the beta-functionals of the open string background fields $\nu$ and $A^\text{\scalebox{0.8}{B}}_\alpha$ to zero, giving rise to the spacetime equations of motion that determine the low-energy dynamics of the D-brane \cite{Gomis:2020fui}. It is shown in \cite{Gomis:2020fui} that these equations of motion arise from a Dirac-Born-Infeld-like (DBI) action
\be \label{eq:dd2a}
    S_{\text{D}(d-2)} = - T_{d-2} \int d^{d-1} Y \, e^{-\Phi} \sqrt{- \det
    \begin{pmatrix}
        0 &\,\,\,\, \p_\beta f^\nu \, \tau_\nu \\[2pt]
        \p_\alpha f^\mu \, \bar{\tau}_\mu &\,\,\,\, \p_\alpha f^\mu \, \p_\beta f^\nu \, E_{\mu\nu} + \CF_{\alpha\beta}
    \end{pmatrix}}\,,
\ee
where $\CF_{\alpha\beta} = \p_\alpha f^\mu \, \p_\beta f^\nu \, B_{\mu\nu} + F_{\alpha\beta}$\,, with $F_{\alpha\beta} = \p_\alpha A^\text{\scalebox{0.8}{B}}_\beta - \p_\beta A^\text{\scalebox{0.8}{B}}_
\alpha$\,. Formally, \eqref{eq:dd2a} looks rather similar to the relativistic case; however, a crucial difference makes this action fundamentally distinct from the standard DBI action: the matrix is $d \times d$ instead of $(d-1) \times (d-1)$ as is the case for relativistic D($d-2$)-branes! This D-brane action can be readily generalized to any D$p$-branes, also in the presence of Ramond-Ramond potentials, with \cite{Ebert:2021mfu}
\be \label{eq:dpaction}
    S_{\text{D}p} = - T_p \int d^{p+1} \sigma \, e^{-\Phi} \sqrt{-\det
    \begin{pmatrix}
        0 &\,\,\,\, \tau_\beta \\[2pt]
        \bar{\tau}_\alpha &\,\,\,\, E_{\alpha\beta} + \CF_{\alpha\beta}
    \end{pmatrix}}
        - T_p \int \sum_{q} C^{(q)} \wedge e^{\CF} \Big|_{p+1}\,,
\ee
where
\be \label{eq:tpdef}
    T_p = \frac{1}{(2\pi)^p \, \bigl( \alpha' \bigr)^{(p+1)/2}}\,.
\ee
It is understood that only $(p+1)$-forms are kept in the Wess-Zumino term. Here, $p$ is even for type IIA and odd for type IIB superstring theory.
We use the basis in which the action \eqref{eq:dpaction} is invariant under the Ramond-Ramond gauge transformations,
\be \label{eq:rrgs}
    \delta^{}_\text{RR} C^{(q)} = d \zeta^{(q-1)} + dB \wedge \zeta^{(q-3)}\,,
\ee
where $\zeta^{(q)} = 0$ if $q \leq 0$\,. The action is also invariant under the Kalb-Ramond two-form transformation,
\be \label{eq:krgs}
    \delta^{}_\text{KR} B = d\xi\,,
        \qquad%
    \delta^{}_\text{KR} A^\text{\scalebox{0.8}{B}} = - \xi\,.
\ee
In flat spacetime,\,\footnote{In the case where the longitudinal spatial direction is compactified over a circle, we fix its radius to be unity, such that $\tau_\mu{}^A = \delta_\mu^A$\,.} the effective nonrelativistic D$p$-brane tension is given by
\be
    T^{}_{\text{D}p} = \frac{T_p}{g_s}\,,
        \qquad\qquad%
    g_s = e^{\langle \Phi \rangle}\,.
\ee
Note that the action \eqref{eq:dpaction} also captures the D-brane configuration that extends in the longitudinal spatial direction, which leads to noncommutative open string theory with a relativistic string spectrum and noncommutativity between space and time \cite{Danielsson:2000mu}.

\section{Nonrelativistic \texorpdfstring{(\emph{p}\,,\,\emph{q})}{(p,q)}-Strings and \texorpdfstring{SL(2,\,$\mathbb{Z}$}{SLZ}) Duality}
\label{sec:nrpq}

In this section, we construct the SL(2,\,$\mathbb{Z}$)-invariant $(p\,,q)$-string action in nonrelativistic string theory and show that it contains two different branches representing strings carrying opposite two-form charges.
However, the strings in either branch do not form any closed group representation individually. {Instead, the two branches transform into each other under the full SL(2,\,$\mathbb{Z}$) group. We will show how this works at the level of the string action and the string states (see after Eq.~\eqref{eq:sm}).} While the background fields transform under the SL(2,\,$\mathbb{Z}$), the boundary between the two branches also rotates accordingly. As we will see below, this implies that the SL(2,\,$\mathbb{Z}$) transformation is realized in a highly non-conventional way in nonrelativistic string theory. In Appendix~\ref{sec:NRlimits}, we will show how this branched phenomenon in nonrelativistic string theory arises from a well-defined limit of relativistic string theory.

\subsection{Nonrelativistic \texorpdfstring{(\emph{p}\,,\,\emph{q})}{pq}-Strings} \label{sec:nrd1b}

It is known that the relativistic ($p\,,q$)-string action can be derived from dualizing the $U(1)$ gauge potential on the D1-brane action in relativistic string theory \cite{Townsend:1997kr, Townsend:1996xj, Aganagic:1997zk}. The same mechanism has also been realized in nonrelativistic string theory \cite{Ebert:2021mfu}, which we review in the following. This analysis gives rise to the nonrelativistic $(p\,,q)$-string action, which enables our analysis of the SL(2,\,$\mathbb{Z}$) duality in nonrelativistic string theory.

From the nonrelativistic D$p$-brane action \eqref{eq:dpaction}, we read off the action for $q$ decoupled nonrelativistic D1-branes
\be \label{eq:d1a}
    S_\text{D1} = - |q| \, T_1 \int d^2 \sigma \, e^{-\Phi} \sqrt{-
    \det
    \begin{pmatrix}
        0 &\,\,\,\, \tau_\beta \\[2pt]
        \bar{\tau}_\alpha &\,\,\,\, E_{\alpha\beta} + \CF_{\alpha\beta}
    \end{pmatrix}}
        - q \, T_1 \int  \lr C^{(2)} + C^{(0)} \, \CF \rr,
\ee
where, for later use, we allow $q$ to be negative. Here, $T_1$ denotes the fundamental string tension. It then follows that the tension for a single D1-string is
\be \label{eq:td1}
    T_\text{D1} = \frac{T_1}{g_s}\,.
\ee
Note that the $q$ in front of the DBI-like term has to appear as an absolute value for the theory to be positive definite.
Next, we perform a worldsheet duality transformation of the one-form gauge field $A_\alpha$ by
introducing a parent action that includes a generating functional as follows:
\be \label{eq:sd1p}
    S_\text{parent} = S_\text{D1} + \frac{T_1}{2} \int d^2 \sigma \, \tilde{H}^{\alpha\beta} \lr F_{\alpha\beta} - 2 \, \p_{[\alpha} A^\text{\scalebox{0.8}{B}}_{\beta]} \rr.
\ee
Here, $F$ is treated as an independent field.
Integrating out the auxiliary anti-symmetric two-tensor $\tilde{H}^{\alpha\beta}$ imposes $F = d A^\text{\scalebox{0.8}{B}}$ and leads back to the original action \eqref{eq:d1a}.
Integrating out $A^\text{\scalebox{0.8}{B}}_\alpha$ imposes the equation $\p_\alpha \tilde{H}^{\alpha\beta} = 0$\,. In two dimensions, this equation can be solved locally by setting $\tilde{H}^{\alpha\beta} = p \, \epsilon^{\alpha\beta}$, where $p$ will turn out to be an integer that counts the number of fundamental strings in the dual theory. Performing the electromagnetic duality transformation by further integrating out $F$\,, we find the dual $(p\,,q)$-string action \cite{Ebert:2021mfu},\,\footnote{The procedure of dualizing the worldsheet $U(1)$ potential only corresponds to an SL(2,\,$\mathbb{Z}$) transformation when $q=\pm1$\,.}
\vspace{-4mm}
\begin{empheq}[box=\fbox]{align} \label{eq:spq}
    \quad \notag \\[-6pt]
    \,
    S^+_\text{string} = - \frac{T_1}{2} \int d^2 \sigma \, \bigl( p - q \, C^{(0)} \bigr)  \sqrt{-\tau} \, \tau^{\alpha\beta} \, E_{\alpha\beta} - T_1 \int \! \lr p \, B + q \, C^{(2)} + \frac{1}{2} \frac{q^2 \, e^{-2\Phi} \, \ell}{p - q \, C^{(0)}} \rr
    \, \\[-10pt]
    \notag
\end{empheq}
Here, $\ell$ is a two-form that is defined in components to be
\begin{align}\label{eq:TSrescalC}
    \ell^{}_{\alpha\beta} = \tau^{}_\alpha{}^A \, \tau^{}_\beta{}^B \, \epsilon_{AB}\,.
\end{align}
Without loss of generality, we assumed that
\be
    \star\ell = \frac{1}{2} \, \epsilon^{\alpha\beta} \, \tau_\alpha{}^A \, \tau_\beta{}^B \, \epsilon_{AB} > 0\,.
\ee
This restriction is consistent with the connected component of the local $\mathrm{SO(1,1)}$ in the two-dimensional longitudinal sector of the target-space geometry.
The $(p\,,q)$-string action has to be supplemented with the condition $p - q \, C^{(0)} > 0$\,. This is because the equation of motion from varying $F$ in Eq.~\eqref{eq:sd1p} gives
\be \label{eq:pqclz}
    p - q \, C^{(0)} = |q| \, e^{-\Phi} \sqrt{- \tau} \ls - \det \begin{pmatrix}
        0 &\,\,\,\, \tau_\beta \\[2pt]
        \bar{\tau}_\alpha &\,\,\,\, E_{\alpha\beta} + \CF_{\alpha\beta}
    \end{pmatrix} \rs^{\!-\frac{1}{2}} \!\! > 0\,.
\ee
Moreover, the condition $p - q \, C^{(0)} > 0$ is necessary for the {kinetic part of the} action \eqref{eq:spq} to be positive definite. For this reason, we labeled the ($p\,,q$)-string action \eqref{eq:spq} with ``$+$" as $S^+_\text{string}$\,. Quantum mechanically, $p$ and $q$ are required to be integers and label the number of fundamental strings and D1-strings in the composite $(p\,,q)$-string state. Note that, for the above procedure to make sense, we have assumed that $q \neq 0$ and $p - q \, C^{(0)} \neq 0$\,. However, when $p \neq 0$ and $q = 0$\,, the action \eqref{eq:spq} still holds and describes $p$ fundamental nonrelativistic strings, in agreement with the Nambu-Goto action \eqref{eq:ngfsa}. This implies that the condition $q \neq 0$ arises only from a shortcoming of the D1-string action \eqref{eq:d1a} instead of any physical limitation, which {can} be avoided by using the Hamiltonian formalism {(see page 47 in \cite{Townsend:1996xj} for an extensive argument)}.


Performing the same duality transformation for the action \eqref{eq:d1a} but with $q$ replaced by $-q$ and substituting the solution $\tilde{H}^{\alpha\beta} = - p \, \epsilon^{\alpha\beta}$\,, we are led to the following action:
\vspace{-4mm}
\begin{empheq}[box=\fbox]{align} \label{eq:spqm}
    \quad \notag \\[-6pt]
    \,
    S^-_\text{string} = \frac{T_1}{2} \int d^2 \sigma \, \bigl( p - q \, C^{(0)} \bigr)  \sqrt{-\tau} \, \tau^{\alpha\beta} \, E_{\alpha\beta} + T_1 \int  \lr p \, B + q \, C^{(2)} + \frac{1}{2} \frac{q^2 \, e^{-2\Phi} \, \ell}{p - q \, C^{(0)}} \rr
    \, \\[-10pt]
    \notag
\end{empheq}
This action is supplemented with the condition
\be
    p - q \, C^{(0)} < 0\,,
\ee
opposite to the condition \eqref{eq:pqclz}.
The string states captured by $S^+_\text{string}$ and $S^-_\text{string}$ carry the opposite $(p,\,q)$ charges, and are related to each other via the transformation $(p\,, q) \rightarrow (-p\,, -q)$\,, leaving the background fields unchanged.
We will see that both $S^+_\text{string}$ and $S^-_\text{string}$ are required to realize the SL(2,\,$\mathbb{Z}$) transformation in nonrelativistic string theory. We will comment on the boundary case with $p - q \, C^{(0)} = 0$ in Section~\ref{sec:wstl} for nonrelativistic $(p\,,q)$-strings and Section~\ref{sec:ibd3b} for nonrelativistic D3-branes.

Generically, it can be a nontrivial matter to extract the effective tension of the $(p\,,q)$-string from the action \eqref{eq:spq} (or Eq.~\eqref{eq:spqm}). The tension is the coefficient of the kinetic term of the Nambu-Goldstone bosons describing the fluctuations of the string. In the following, we compute the effective tension in the special case when the $(p\,,q)$-string extends in the longitudinal spatial direction. In flat spacetime with $\tau_\mu{}^A = \delta_\mu^A$ and $E_\mu{}^{A'} = \delta_\mu^{A'}$\,, the string configuration is described by
\begin{align} \label{eq:bconf}
    X^0 = \sigma^0\,,
        \qquad%
    X^1 = \sigma^1\,,
        \qquad%
    X^{A'} = X_0^{A'} + \pi^{A'} ( \sigma^\alpha )\,,
\end{align}
where the $(p\,,q)$-string is localized at the transverse directions $X_0^{A'}$\,. The fields $\pi^{A'}$ denote the associated Nambu-Goldstone bosons that arise from spontaneously breaking the translational symmetries, and they describe the fluctuations of the shape of the $(p\,,q)$-string. Expanding the action \eqref{eq:spq} (or Eq.~\eqref{eq:spqm}) to quadratic order in $\pi^{A'}$, we find,
\be \label{eq:etpqnr}
    S^\pm_\text{string} \sim \mp \frac{T_1}{2} \int d^2 \sigma \, \bigl( p - q \, C^{(0)} \bigr) \, \p^\alpha \pi^{A'} \, \p_\alpha \pi^{A'} + \cdots
\ee
In this special case, the quantity
\be \label{eq:nrstension}
    T_{p,\,q} = T_1 \, \bigl| p - q \, \langle C^{(0)} \bigr\rangle \bigr|
\ee
plays the role of the effective tension of the $(p\,, q)$-string. Such $(p\,,q)$-string tensions satisfy the following triangle inequality:
\be \label{eq:tine}
    T_{p_1 + p_2,\, q_1 + q_2} \leq T_{p_1, \,q_1} + T_{p_2, \,q_2}\,.
\ee
This inequality is saturated if all the $(p\,,q)$-strings involved in Eq.~\eqref{eq:tine} fall in the same branch of $(p\,,q)$-space. The $(p\,,q)$-strings saturating the inequality are marginally stable bound states \cite{Dabholkar:1989jt}.

\subsection{A Branched \texorpdfstring{SL(2,\,$\mathbb{Z}$)}{SLZ} Duality} \label{sec:bslzd}

Now, we are ready to study the SL(2,\,$\mathbb{Z}$) transformations of the background fields and analyze how these transformations act on the nonrelativistic $(p\,,q)$-string actions \eqref{eq:spq} and \eqref{eq:spqm} in different branches. This analysis will uncover a branched structure of the SL(2,\,$\mathbb{Z}$) duality in nonrelativistic string theory.

\subsubsection*{$\bullet$ The positive branch}

To observe how the nonrelativistic $(p\,,q)$-string transforms under the SL(2,$\mathbb{Z}$) group, it is useful to introduce variables in the Einstein frame, with
\be
    \bigl( \tau^\text{\scalebox{0.8}{E}}_{\alpha\beta}\,, E^\text{\scalebox{0.8}{E}}_{\alpha\beta}\,, \ell^\text{\scalebox{0.8}{E}} \bigr) = e^{-\Phi/2} \, \bigl( \tau_{\alpha\beta}\,, E_{\alpha\beta}\,, \ell \bigr)\,,
\ee
We then rewrite the action \eqref{eq:spq} as
\be \label{eq:spqrw}
    S^+_\text{string} = - \frac{T_1}{2} \int d^2 \sigma \, \bigl( \Theta^\intercal \, \CC \bigr) \, \sqrt{-\tau^\text{\scalebox{0.8}{E}}} \, \tau^{\alpha\beta}_\text{\scalebox{0.8}{E}} \, E^\text{\scalebox{0.8}{E}}_{\alpha\beta} - T_1 \int  \Theta^\intercal \,   \Sigma\,,
\ee
where
\be \label{eq:tcs}
     \Theta =
    \begin{pmatrix}
        p \\[2pt]
        q
    \end{pmatrix},
        \qquad%
    \CC = e^{\Phi/2}
    \begin{pmatrix}
        1 \\[2pt]
        - C^{(0)}
    \end{pmatrix},
        \qquad%
    \Sigma =
    \begin{pmatrix}
        B \\[2pt]
        C^{(2)}
    \end{pmatrix}
    +
    \frac{q}{p - q \, C^{(0)}}
    \begin{pmatrix}
        0 \\[2pt]
        \frac{1}{2} \, \ell \, e^{-2\Phi}
    \end{pmatrix}.
\ee
We require that the vielbeine be SL(2,$\mathbb{Z}$) invariant in the Einstein frame.
Here, we assume that $(p, q)$ forms a doublet under the \SLZ group and transforms as
\be \label{eq:sl2z1}
   \Theta' = \Lambda \, \Theta\,,
\ee
where
\be\label{eq:sl2zpar}
    \Lambda
        =
    \begin{pmatrix}
        a &\,\, b \\[2pt]
        c &\,\, d
    \end{pmatrix}\,,
\ee
Here, $a, \, b, \, c, \, d \in \mathbb{Z}$ are SL(2,\,$\mathbb{Z}$) group parameters satisfying $a \, d - b \, c = 1$\,. For the first term in \eqref{eq:spqrw} to be SL(2,\,$\mathbb{Z}$) invariant, $\CC$ has to be a doublet as well and thus transforms as
\be \label{eq:sl2z2}
    \CC' = \bigl( \Lambda^{-1} \bigr)^\intercal \, \CC\,,
\ee
which further implies that
\be \label{eq:c0phitrans}
    C'{}^{(0)} = \frac{a \, C^{(0)} + b}{c \, C^{(0)} + d}\,,
        \qquad%
    e^{\Phi'/2} = \bigl( c \, C^{(0)} + d \bigr) \, e^{\Phi/2}\,.
\ee
This transformation only makes sense when $c \, C^{(0)} + d > 0$\,. This is also the requirement that preserves the condition $p - q \, C^{(0)} > 0$\,: under SL(2,\,$\mathbb{Z}$), we have
\begin{equation} \label{eq:pqc0t}
    p' - q' \, C'{}^{(0)} = \frac{p - q \, \Cz}{c \, \Cz + d} > 0\,.
\end{equation}
It looks like only the SL(2,\,$\mathbb{Z}$) transformations that preserve the condition $p - q \, C^{(0)} > 0$ survive. Note that $C^{(0)}$ also transforms nontrivially as in Eq.~\eqref{eq:c0phitrans}.

An obvious question is: where does the other half of the SL(2,\,$\mathbb{Z}$) transformations satisfying the condition $c \, C^{(0)} + d < 0$ go?\,\footnote{When $C^{(0)}$ is rational, it is also possible that $c \, C^{(0)} + d = 0$\,. We exclude this singular case.} Before answering this question, let us first stick to the branch satisfying $c \, C^{(0)} + d > 0$ and understand how the two-form fields $B$ and $C^{(2)}$ transform under the action of SL(2,\,$\mathbb{Z}$), such that the second term in Eq.~\eqref{eq:spqrw} is invariant. Na\"{i}vely, one would expect that the quantity $\Sigma$ in Eq.~\eqref{eq:tcs} has to transform as a doublet, with $\Sigma' = (\Lambda^{-1})^\intercal \, \Sigma$\,, for the term $\Theta^\intercal \, \Sigma$ to be SL(2,\,$\mathbb{Z}$) invariant. However, the definition of $\Sigma$ in Eq.~\eqref{eq:tcs} contains a $(p\,,q)$-dependent term, and forcing $\Sigma$ to be a doublet would require that SL(2,\,$\mathbb{Z}$) act on both $B$ and $C^{(2)}$ in a $(p\,,q)$-dependent way. Such a $(p\,,q)$-dependent transformation {of the supergravity fields $B$ and $C^{(2)}$} would not form a symmetry of type IIB nonrelativistic supergravity: the background fields would have to transform differently depending on how many fundamental and D1-strings are present, even in the case where no back reaction to the background fields is included!

Therefore, {since the quantity $
\Sigma$ is not a supergravity field, it makes perfect sense that we find that $\Sigma$ is a ``quasi"-doublet, meaning that it transforms as a doublet up to a $(p\,,q)$-dependent term,} \emph{i.e.},
\be \label{eq:spcm}
    \Sigma' = \bigl( \Lambda^{-1} \bigr)^\intercal \lr \Sigma + \alpha \, \Theta^\perp \rr,
        \qquad%
    \Theta^\perp =
    \begin{pmatrix}
        q \\[2pt]
        -p
    \end{pmatrix}.
\ee
This already guarantees that the combination $\Theta^\intercal \, \Sigma$ is invariant.
Requiring that both the transformed background fields $B'$ and $C'{}^{(2)}$ be independent of $p$ and $q$, we find a unique solution to $\alpha$\,, with
\be \label{eq:ava}
    \alpha = \frac{c \, \ell}{2 \, e^{2 \Phi} \, \bigl( c \, C^{(0)} + d \big) \, \bigl( p - q \, C^{(0)} \bigr)}\,.
\ee
Together with Eq.~\eqref{eq:c0phitrans}, we find a well-defined set of SL(2,$\mathbb{Z}$) transformations of the background fields $B$\,, $C^{(0)}$, $C^{(2)}$ and $\Phi$\,, given by
\begin{subequations} \label{eq:sl2t}
\begin{align}
    C'{}^{(0)} & = \frac{a \, C^{(0)} + b}{c \, C^{(0)} + d}\,,
        \qquad%
    \Phi' = \Phi + 2 \, \ln \bigl( c \, C^{(0)} + d \bigr)\,, \\[4pt]
    \begin{pmatrix}
        B' \\[2pt]
        C'{}^{(2)}
    \end{pmatrix}
    & =
    \bigl( \Lambda^{-1} \bigr)^\intercal \ls
    \begin{pmatrix}
        B \\[2pt]
        C^{(2)}
    \end{pmatrix} - \frac{c}{c \, C^{(0)} + d} \begin{pmatrix}
        0 \\[2pt]
        \frac{1}{2} \, \ell \, e^{-2\Phi}
    \end{pmatrix} \rs
    - \frac{c}{\bigl( c \, C^{(0)} + d \bigr)^2}
    \begin{pmatrix}
        0 \\[2pt]
        \frac{1}{2} \, \ell \, e^{-2\Phi}
    \end{pmatrix},
\end{align}
\end{subequations}
which are independent of $p$ and $q$\,. These transformations have to satisfy $c \, C^{(0)} + d > 0$ for the ($p\,,q$)-string action \eqref{eq:spq} to be invariant. {In Appendix~\ref{sec:NRlimits}, we show that the same transformations in Eq.~\eqref{eq:sl2t} can be obtained equivalently by taking a well-defined limit of the SL(2,\,$\mathbb{Z}$) transformations in relativistic string theory.}

\subsubsection*{$\bullet$ The negative branch}

As we have discussed earlier, the transformations \eqref{eq:sl2t} with the condition $c \, C^{(0)} + d > 0$ does not yet constitute the full group of SL(2,\,$\mathbb{Z}$)\,. What if we apply the same transformations \eqref{eq:sl2t} to the ($p\,,q$)-string action \eqref{eq:spq}, but now with the condition $c \, C^{(0)} + d < 0$\,? This procedure maps the action $S^+_\text{string}$ in Eq.~\eqref{eq:spqrw} to $S^-_\text{string}$ in Eq.~\eqref{eq:spqm}, \emph{i.e.},
\begin{figure}[t!]
\centering
\begin{tikzpicture}
    \draw (0,0) node (a) {\scalebox{0.35}{\includegraphics[width=\textwidth]{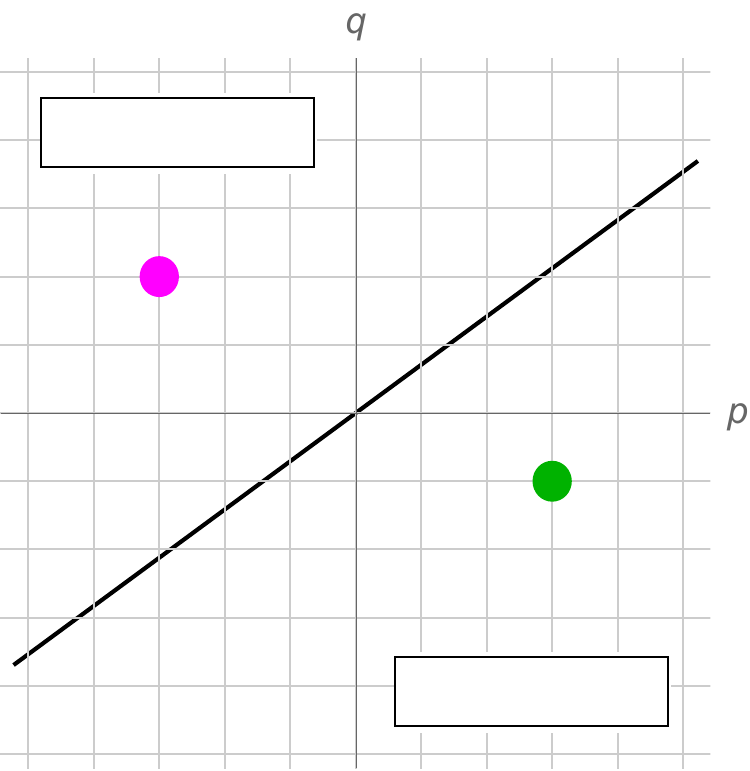}}} (9.5,3) node (b) {\scalebox{0.35}{\includegraphics[width=\textwidth]{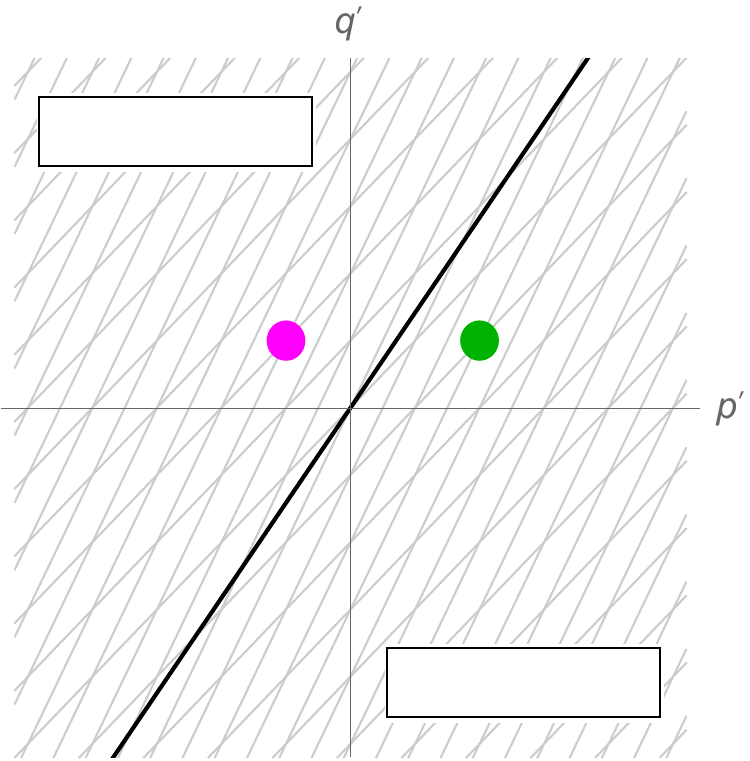}}};
    \draw (0,0) node (a) {\scalebox{0.35}{\includegraphics[width=\textwidth]{figa1.pdf}}} (9.5,-3) node (c) {\scalebox{0.35}{\includegraphics[width=\textwidth]{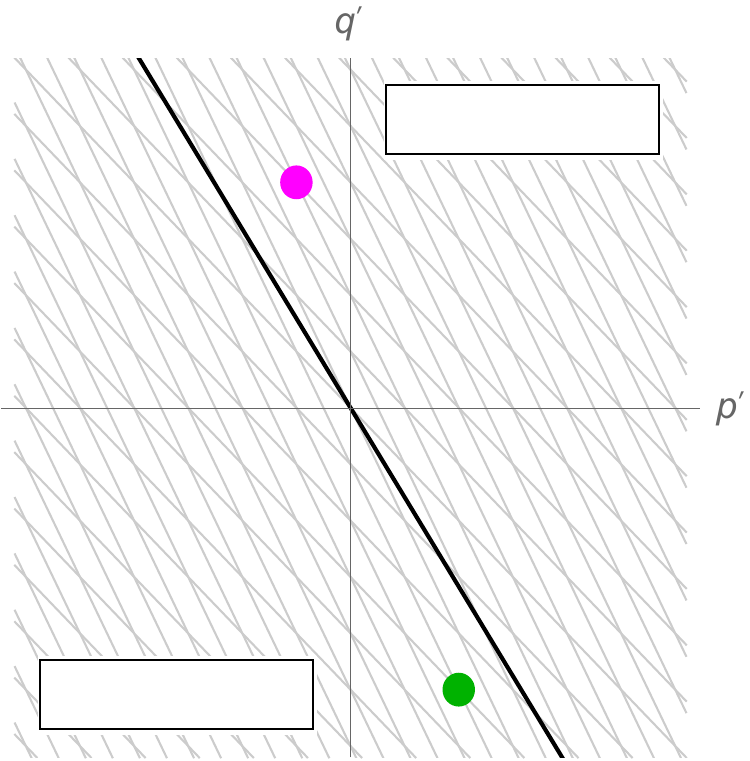}}};
    \draw[->, thick] (a)->(b);
    \draw[->, thick] (a)->(c);
    \draw (5-0.5,1.8) node {\rotatebox{15}{\scalebox{0.9}{$c \, C^{(0)} + d > 0$}}};
    \draw (5-0.5,-1.8) node {\rotatebox{-15}{\scalebox{0.9}{$c \, C^{(0)} + d < 0$}}};
    \draw (5.2-0.4,0) node {\textbf{SL(2,\,$\mathbb{Z}$)}};
    \draw (-1.43,1.88) node {\scalebox{0.7}{$p \, - \, q \,\, {C}^{(0)} \, < \, 0$}};
    \draw (1.15,-2.2) node {\scalebox{0.7}{$p \, - \, q \,\, {C}^{(0)} \, > \, 0$}};
    \draw (8.56-0.5,4.83) node {\scalebox{0.7}{$p' - q' \, {C'}^{(0)} < 0$}};
    \draw (11.08-0.5,0.83) node {\scalebox{0.7}{$p' - q' \, {C'}^{(0)} > 0$}};
    \draw (8.56-0.5,-5.25) node {\scalebox{0.7}{$p' - q' \, {C'}^{(0)} < 0$}};
    \draw (11.08-0.5,-1.068) node {\scalebox{0.7}{$p' - q' \, {C'}^{(0)} > 0$}};
\end{tikzpicture}
\caption{The SL(2,\,$\mathbb{Z}$) action in ($p\,, q$)-space. We marked one point in each branch and plotted the location to which they are mapped under an SL(2,\,$\mathbb{Z}$) transformation satisfying $c \, C^{(0)} + d > 0$ and another one satisfying $c \, C^{(0)} + d < 0$\,. In the first case, the colored points remain in the same branch, whereas in the second, they switch branches. The grids are aids to visualize how the branched SL(2,\,$\mathbb{Z}$) transformation acts on the plane.}
\label{fig:comoving}
\end{figure}
\be \label{eq:sm}
    S^-_\text{string} = \frac{T_1}{2} \int d^2 \sigma \, \bigl( \Theta^\intercal \, \CC \bigr) \, \sqrt{-\tau^\text{\scalebox{0.8}{E}}} \, \tau^{\alpha\beta}_\text{\scalebox{0.8}{E}} \, E^\text{\scalebox{0.8}{E}}_{\alpha\beta} + T_1 \int \Theta^\intercal \,  \Sigma\,,
\ee
where we performed a discrete transformation $B' \rightarrow - B'$ and $C'{}^{(2)} \rightarrow - C'{}^{(2)}$ and dropped the primes that label the \SLZ-dual fields in the result.
In this dual frame, according to the transformation in Eq.~\eqref{eq:pqc0t}, we are now in the branch of $p - q \, C^{(0)} < 0$\,, which is precisely the condition for {the kinetic part of} $S^-_\text{string}$ to be positive definite. The SL(2,\,$\mathbb{Z}$) transformations that keep $S^-_\text{string}$ invariant are the same as the ones given in Eq.~\eqref{eq:sl2t}.
We therefore conclude the following:
\begin{itemize}

\item

the SL(2,\,$\mathbb{Z}$) transformations satisfying $c \, C^{(0)} + d > 0$ preserve the string action $S^+_\text{string}$ ($S^-_\text{string}$) in the positive (negative) $p - q \, C^{(0)}$ branch in the space of ($p\,,q$)-string states;

\item

the SL(2,\,$\mathbb{Z}$) transformations satisfying $c \, C^{(0)} + d < 0$ map between $S^+_\text{string}$ and $S^-_\text{string}$\,.

\end{itemize}

\vspace{3mm}

Therefore, the SL(2,\,$\mathbb{Z}$) duality in type IIB nonrelativistic superstring theory relates two different branches representing nonrelativistic ($p\,, q$)-string states satisfying $p - q \, C^{(0)} > 0$ and the states satisfying $p - q \, C^{(0)} < 0$\,, respectively. These two branches are separated by the line $p - q \, C^{(0)} = 0$ in the ($p\,, q$)-plane and they are described by the ($p\,, q$)-string actions $S^{+}_\text{string}$ and $S^-_\text{string}$\,, respectively. The boundary $p - q \, C^{(0)} = 0$ itself changes with respect to the SL(2,\,$\mathbb{Z}$) transformations as $C^{(0)}$ transforms nontrivially. The SL(2,\,$\mathbb{Z}$) transformations are also branched: the transformations satisfying $c \, C^{(0)} + d > 0$ map between string states in the same branch, while the transformations satisfying $c \, C^{(0)} + d < 0$ map between different branches. See Figure~\ref{fig:comoving} for a summary of the SL(2,\,$\mathbb{Z}$) action in $(p\,,q)$-space.

\subsection{Manifestly \texorpdfstring{SL(2,\,$\mathbb{Z}$)}{Z} Invariant String Action} \label{sec:mslia}

Collectively, the above branched realization of SL(2,\,$\mathbb{Z}$) invariance can be summarized as follows: The invariant action that unifies $S^\pm_\text{string}$ is
\be \label{eq:ta}
    S_\text{string} = - \frac{T^{}_1}{2} \int d^2 \sigma \, \bigl| \Theta^\intercal \, \CC \bigr| \, \sqrt{-\tau^\text{\scalebox{0.8}{E}}} \, \tau^{\alpha\beta}_\text{\scalebox{0.8}{E}} \, E^\text{\scalebox{0.8}{E}}_{\alpha\beta} - T^{}_1 \int \bigl( \Theta^\intercal \,   \Sigma \bigr)\, \text{sgn} \bigl( p - q \, C^{(0)} \bigr)\,.
\ee
The SL(2,\,$\mathbb{Z}$) transformations are
\be \label{eq:tc}
    \Theta' = \Lambda \, \Theta\,,
        \qquad%
    \CC' = \text{sgn} (c \, C^{(0)} + d) \, \bigl( \Lambda^{-1} \bigr)^\intercal \, \CC\,,
\ee
and
\be \label{eq:spsgn}
    \Sigma' = \sgn (c \, C^{(0)} + d) \, \bigl( \Lambda^{-1} \bigr)^\intercal \ls \Sigma + \frac{c \, \ell}{2 \, e^{2 \Phi} \, \bigl( c \, C^{(0)} + d \big) \, \bigl( p - q \, C^{(0)} \bigr)} \, \Theta^\perp \rs.
\ee
The variables $\bigl( \tau^\text{\scalebox{0.8}{E}}_{\alpha\beta}\,, E^\text{\scalebox{0.8}{E}}_{\alpha\beta}\,, \ell^\text{\scalebox{0.8}{E}} \bigr)$ in Einstein's frame are invariant.
Recall that the quantities $\Theta$\,, $\CC$ and $\Sigma$ are defined in Eq.~\eqref{eq:tcs}. Moreover, as defined in Eq.~\eqref{eq:spcm}, $\Theta^\perp = (q\,, -p)^\intercal$\,.
The SL(2,\,$\mathbb{Z}$) invariance of Eq.~\eqref{eq:ta} is made manifest by noticing that
\be
    \text{sgn} \bigl( p' - q' \, C'{}^{(0)} \bigr) = \text{sgn} \! \lr \frac{p - q \, C^{(0)}}{c \, C^{(0)} + d} \rr.
\ee
Note that the quasi-doublet $\Sigma$ does not only transform as a doublet up to a term orthogonal to $\Theta$\,, but also up to an overall sign, depending on the value of the quantity $c \, C^{(0)} + d$\,.
In components, we find the complete SL(2,\,$\mathbb{Z}$) transformations that generalize Eq.~\eqref{eq:sl2t},
\begin{subequations} \label{eq:slztfs}
\begin{align}
    C'{}^{(0)} & = \frac{a \, C^{(0)} + b}{c \, C^{(0)} + d}\,,
        \qquad%
    \Phi' = \Phi + 2 \, \ln |c \, C^{(0)} + d|\,, \\[4pt]
    \begin{pmatrix}
        B' \\[2pt]
        C'{}^{(2)}
    \end{pmatrix}
    & = \text{sgn} \bigl( c \, C^{(0)} + d \bigr) \,
    \Bigg\{ \bigl( \Lambda^{-1} \bigr)^\intercal \ls
    \begin{pmatrix}
        B \\[2pt]
        C^{(2)}
    \end{pmatrix} - \frac{c}{c \, C^{(0)} + d} \begin{pmatrix}
        0 \\[2pt]
        \frac{1}{2} \, \ell \, e^{-2\Phi}
    \end{pmatrix} \rs \notag
    \\[2pt]
    & \hspace{5.7cm} - \frac{c}{\bigl( c \, C^{(0)} + d \bigr)^2}
    \begin{pmatrix}
        0 \\[2pt]
        \frac{1}{2} \, \ell \, e^{-2\Phi}
    \end{pmatrix} \Bigg\}\,. \label{eq:bctf}
\end{align}
\end{subequations}
These transformations will later be extended to include the one acting on the four-form Ramond-Ramond potential $C^{(4)}$ in Eq.~\eqref{eq:c4trnsf}.

\subsection{Winding Strings in a Tensionless Limit} \label{sec:wstl}

So far, we only considered nonrelativistic ($p\,,q$)-strings satisfying the condition $p - q \,C^{(0)} \neq 0$\,. In this last subsection, we consider a class of inter-branch strings at $p - q \,C^{(0)} = 0$\,.
One such $p - q \, C^{(0)} \rightarrow 0$ limit arises if the tension $T_1$ is simultaneously scaled to zero, while keeping finite the \SLZ-invariant quantity
\be
    T^{}_\text{eff} \equiv \lim_{\substack{\\
    T_1 \rightarrow 0 \\[2pt]
    C^{(0)} \rightarrow \frac{p}{q}}} T^{}_1 \, e^{-\Phi} \, \biggl| \frac{q}{p -q \, C^{(0)}} \biggr|\,.
\ee
The \SLZ-invariance of such an effective tension $T_\text{eff}$ at {$p - q \, C^{(0)} = 0$} follows directly from the SL(2,\,$\mathbb{Z}$) transformations \eqref{eq:sl2z1} and \eqref{eq:sl2t} in Section~\ref{sec:mslia}. Since $T_1 = (2\pi\alpha')^{-1}$ (see Eq.~\eqref{eq:tpdef}), we are essentially taking an infinite Regge slope limit. Under this double scaling limit, the resulting action is
\be \label{eq:tls}
    S_\text{tensionless} = \frac{T_\text{eff}}{2} \int |q| \, e^{-\Phi} \, \ell\,.
\ee
This theory is \SLZ-invariant and purely topological. {Just as in the previous analysis, this action splits into two branches depending on the sign of $q$\,. As before, each branch is mapped into itself under an SL(2,\,$\mathbb{Z}$)  transformation satisfying the condition $c \, \Cz + d > 0$\,, and one branch is mapped into the other branch and vice versa if $c \, \Cz + d < 0$.}

In order to decode the physics described by the new action \eqref{eq:tls}, we consider a simple string configuration in a constant dilaton background with $g_s = e^{\Phi}$ and in the flat geometrical background fields with $\tau_\mu{}^A = \delta_\mu^A$ and $E_\mu{}^{A'} = \delta_\mu^{A'}$\,. We focus on the winding $(p\,, q)$-string configuration defined by
\begin{align}
    X^0 = \sigma^0\,,
        \qquad%
    X^1 = w \, R \, \sigma^1\,,
        \qquad%
    X^{A'} = X_0^{A'} + \pi^{A'} ( \sigma^\alpha )\,,
\end{align}
where $X^1$ is compactified over a circle of radius $R$ and the $(p\,, q)$-string winds $w$ times around $X^1$\,, with $w \in \mathbb{Z}$\,.
In this case, the Lagrangian density of the action \eqref{eq:tls} evaluates to
\be \label{eq:swn}
    \CL_\text{tensionless} = w R \, \frac{|q| \, T_\text{eff}}{g_s}\,,
\ee
which is proportional to how many times the $(p\,,q)$-string winds around the $X^1$ direction.

\section{\texorpdfstring{SL(2,\,$\mathbb{Z}$)}{Z} Symmetric Actions for D-Instanton and D3-Brane} \label{sec:d3}

We have discussed the SL(2,\,$\mathbb{Z}$) transformations of various background fields in type IIB nonrelativistic supergravity, using the $(p\,,q)$-string as a probe. However, the consideration of the $(p\,,q)$-strings does not suffice for establishing the complete SL(2,\,$\mathbb{Z}$) duality in type IIB nonrelativistic supergravity, which also contains a Ramond-Ramond potential $C^{(4)}$. This four-form gauge field is not visible to the $(p\,,q)$-string, but it is coupled to D3-branes. In this section, we construct a manifestly SL(2,\,$\mathbb{Z}$) invariant D3-brane action in nonrelativistic string theory. In Appendix~\ref{sec:NRlimits}, we will show how the action arises from its relativistic counterpart. This formalism will fortify the branched SL$(2,\,\mathbb{Z})$ realization in nonrelativistic string theory,
which we have already learned about from nonrelativistic $(p\,,q)$-strings. Moreover, it will
grant us access to studying the inter-branch states satisfying $p - q \, C^{(0)} = 0$\,.

Just as in the relativistic case, {in order to formulate a manifestly SL(2,\,$\mathbb{Z}$)-symmetric D3-brane action, it is convenient to introduce another doublet} \cite{Bergshoeff:2006gs}
\be \label{eq:tth}
    \tilde{\Theta} =
    \begin{pmatrix}
        \tilde{p} \, \\[2pt]
        \tilde{q}
    \end{pmatrix}
\ee
that is conjugate to $\Theta = (p\,\,\,q)^\intercal$\,, satisfying the condition
\be \label{eq:pqcond1}
    p \, \tilde{q} - q \, \tilde{p} = 1\,.
\ee
Note that $\tilde{\Theta}$ is defined up to an arbitrary shift in $\Theta$\,, forming the equivalence class
\be
    [\tilde{\Theta}]:
        \quad%
    \tilde{\Theta} \sim \tilde{\Theta} + \beta \, \Theta\,,
        \quad%
    \beta \in \mathbb{R}\,.
\ee
We have chosen the normalization of $\tilde{\Theta}$ such that the expression in Eq.~\eqref{eq:pqcond1} is unity.\,\footnote{When $p \, \tilde{q} - q \, \tilde{p} = 0$\,, we have $\tilde{\Theta} \propto \Theta$ and such a $\tilde{\Theta}$ does not constitute a linearly independent doublet.}
Just as $\Theta$\,, the new quantity $\tilde{\Theta}$ transforms under the SL(2,\,$\mathbb{Z}$) symmetry as
\be \label{eq:tthsl2z}
    \tilde{\Theta}' = \Lambda \, \tilde{\Theta}\,.
\ee
The condition \eqref{eq:pqcond1} is manifestly SL(2,\,$\mathbb{Z}$) invariant.

In the special case where $(p\,, q) = (1, 0)$\,, \emph{i.e.}, the $(p\,, q)$-string reduces to a single fundamental string, a conjugate state is the single D-string state with $(\tilde{p}\,, \tilde{q}) = (0, 1)$\,. In this case, $\Theta$ and $\tilde{\Theta}$ are projections to the fundamental and D-string state, respectively.

The new ingredient $\tilde{\Theta}$ is already required for the construction of a ($p\,, q$)-symmetric action for the nonrelativistic D(-1)-brane, which represents an instanton state. We will study this D-instanton first, from which useful ingredients for building a ($p\,, q$)-symmetric nonrelativistic D3-brane action will be obtained.

\subsection{Nonrelativistic D-Instanton} \label{sec:nri}

The nonrelativistic D-instanton action can be read off from the general D$p$-brane action \eqref{eq:dpaction} by setting $p = -1$\,, which gives
\be \label{eq:sd-1}
    S_{\text{D(-1)}} = - T_{-1} \, C^{(0)}\,.
\ee
This action is not manifestly invariant under the SL(2,\,$\mathbb{Z}$) transformation \eqref{eq:sl2t}, which acts nontrivially on $C^{(0)}$. In order to achieve the SL(2,\,$\mathbb{Z}$) symmetry, Eq.~\eqref{eq:sd-1} has to be covariantized as follows:
\begin{align} \label{eq:sld-10}
    S_{\text{D(-1)}}
    = T_{-1} \, \frac{\tilde{\Theta}^\intercal \, \CC}{{{\Theta}}^\intercal \, \CC}\,.
\end{align}
Since $\CC$ defined in Eq,~\eqref{eq:tcs} is an SL(2,\,$\mathbb{Z}$) doublet up to $\sgn (c \, C^{(0)} + d)$\,, the above action is manifestly SL(2,\,$\mathbb{Z}$) invariant. More explicitly, the instanton action \eqref{eq:sld-10} gives\,\footnote{Note that $T_{-1} = 2\pi$ (see Eq.~\eqref{eq:tpdef}) and thus there is \emph{no} analog of the tensionless $T_1$ limit as in Section~\ref{sec:wstl}.}
\begin{align} \label{eq:sld-1}
    \boxed{S_{\text{D(-1)}}
     = T_{-1} \, \frac{\tilde{p} - \tilde{q} \, C^{(0)}}{{p} - {q} \, C^{(0)}}\,.}
\end{align}
In the case where $\Theta = (1\,\,0)^\intercal$ and $\tilde{\Theta} = (0\,\,1)^\intercal$\,, the D-instanton action \eqref{eq:sd-1} is recovered. {Note that the action becomes independent of $\tilde{\Theta}$ after the condition \eqref{eq:pqcond1} is imposed, except for an overall constant shift of the action.}

\subsection{Nonrelativistic D3-Brane} \label{sec:nd3b}

We are finally ready to construct a manifestly \SLZ-invariant nonrelativistic  D3-brane action. Setting $p=3$ in the general nonrelativistic  D$p$-brane action \eqref{eq:dpaction}, and rewriting the geometric data in terms of the Einstein-frame variables, we find
\begin{align} \label{eq:dpactiond3}
\begin{split}
    S_{\text{D}3}
    = & - T_3 \int d^4 \sigma \, \sqrt{-\det
    \begin{pmatrix}
        0 &\,\,\,\, \tau^\text{\scalebox{0.8}{E}}_\beta \\[2pt]
        \bar{\tau}^\text{\scalebox{0.8}{E}}_\alpha &\,\,\,\, E^\text{\scalebox{0.8}{E}}_{\alpha\beta} + e^{-\Phi/2} \, \CF_{\alpha\beta}
    \end{pmatrix}} \\[2pt]
    & - T_3 \int \lr C^{(4)} + C^{(2)} \wedge \CF + \frac{1}{2} \, C^{(0)} \, \CF \wedge \CF  \rr,
\end{split}
\end{align}
where $\CF = B + dA^{\text{\scalebox{0.8}{B}}}$\,.
This action is invariant under the Ramond-Ramond and Kalb-Ramond gauge symmetries given in Eqs.~\eqref{eq:rrgs} and \eqref{eq:krgs}.

We would like to write down a manifestly \SLZ-invariant D3-brane action by replacing the quantities in the action \eqref{eq:dpactiond3} with their \SLZ-invariant counterparts.\,\footnote{See Appendix~\ref{app:ivs} for further justification of the following replacements.} The invariant form of $C^{(0)}$ has already been given in Section~\ref{sec:nri}, where we essentially performed the following replacement:
\be \label{eq:rpc0}
    C^{(0)} \rightarrow - \frac{\tilde{\Theta}^\intercal \, \CC}{\Theta^\intercal \, \CC}\,,
\ee
which reduces to $C^{(0)}$ when $\Theta = (1\,\,0)^\intercal$ and $\tilde{\Theta} = (0\,\,1)^\intercal$\,. Similarly, the \SLZ-invariant counterpart of the dilaton term $e^\Phi$ is given by
\be \label{eq:rpphi}
    e^{\Phi} \rightarrow \bigl( \Theta^\intercal \, \CC \bigr)^2\,.
\ee
We also learned in Section~\ref{sec:nrpq} that the \SLZ-invariant version of the $B$-field is given by Eq.~\eqref{eq:ta}, with
\be \label{eq:brep}
    B \rightarrow \sgn (p - q \, C^{(0)}) \, \bigl(\Theta^\intercal \, \Sigma\bigr)\,,
\ee
where $\Sigma$ is defined in Eq.~\eqref{eq:tcs}.
This implies that the quantity $\CF$ should be replaced as
\be \label{eq:rpf}
    \CF \rightarrow \sgn \bigl(p - q \, C^{(0)}\bigr) \, \bigl( \Theta^\intercal \, \mathscr{F} \bigr)\,,
        \qquad%
    \mathscr{F} = \Sigma + d\CA\,,
\ee
where $\CA = \bigl( A^{\text{\scalebox{0.8}{B}}} \,\, A^{\text{\scalebox{0.8}{C}}} \bigr)^\intercal$\,, with $A^{\text{\scalebox{0.8}{B}}}$ and $A^{\text{\scalebox{0.8}{C}}}$ the Born-Infeld potentials associated with $B$ and $C^{(2)}$, respectively. The Born-Infeld vector $\CA$ transforms under the combined Ramond-Ramond and Kalb-Ramond gauge symmetries as
\be
    \delta \CA = - \Xi\,,
        \qquad%
    \Xi \equiv
    \begin{pmatrix}
        \xi \\[2pt]
        \zeta^{(1)}
    \end{pmatrix},
\ee
such that $\mathscr{F}$ is gauge invariant. We require that $\CA$ be an SL(2,\,$\mathbb{Z}$) doublet up to a sign, satisfying the following transformation:
\be
    \CA' = \sgn \bigl(c \, C^{(0)} + d \bigr) \, (\Lambda^{-1})^\intercal \, \CA\,.
\ee
The introduction of the extra field $A^\text{\scalebox{0.8}{C}}$ does not change the degrees of freedom as long as $\CA$ is only contracted with $\Theta$ (but not $\tilde{\Theta}$) \cite{Bergshoeff:2006gs}.

We still need to construct an \SLZ-invariant counterpart of $C^{(2)}$\,, which has to involve the doublet $\tilde{\Theta}$\,. This requires building a new quasi-doublet $\tilde{\Sigma}$ that transforms as a doublet up to a $(\tilde{p}\,, \tilde{q})$-dependent term. In analogy with the quantity $\Sigma$ in Eq.~\eqref{eq:spsgn}, we have
\be \label{eq:tspqd}
    \tilde{\Sigma}' = \sgn \bigl( c \,  C^{(0)} + d \bigr) \, \bigl( \Lambda^{-1} \bigr)^\intercal \, \Bigl( \tilde{\Sigma} + \tilde{\alpha} \, \tilde{\Theta}^\perp \Bigr)\,,
        \qquad%
    \tilde{\Theta}{}^\perp_{\phantom{\dagger}} =
    \begin{pmatrix}
        \tilde{q} \,\\[2pt]
        - \tilde{p}
    \end{pmatrix}\,.
\ee
We have introduced $\tilde{\Theta}{}^\perp_{\phantom{\dagger}}$ that is orthogonal to $\tilde{\Theta}$\,, satisfying $\tilde{\Theta}{}^\intercal_{\phantom{\dagger}} \, \tilde{\Theta}{}^\perp_{\phantom{\dagger}} = 0$\,.
Such a quasi-doublet $\tilde{\Sigma}$ defined with respect to $\tilde{\Theta}$ is derived in Appendix \ref{app:ivs}, with
\be \label{eq:tsdef}
    \tilde{\Sigma} = \Sigma + \frac{q \, \tilde{q}{\,}^{-1}}{\bigl( p - q \, C^{(0)} \bigr)^2}
    \begin{pmatrix}
        0 \\[2pt]
        \frac{1}{2} \, \ell \, e^{-2\Phi}
    \end{pmatrix}\,,
\ee
from which we find that the parameter $\tilde{\alpha}$ in Eq.~\eqref{eq:tspqd} is given by
\be
    \tilde{\alpha} = \frac{c \, e^{-2\Phi} \, \ell}{2 \, \bigl( c \, C^{(0)} + d \big) \, \bigl( p - q \, C^{(0)} \bigr) \, \bigl( c \, \tilde{p} + d \, \tilde{q} \, \bigr)} \lr c \, p + d \, q + \frac{q}{\tilde{q}} \, \frac{c \, C^{(0)} + d}{p - q \, C^{(0)}} \rr.
\ee
The appropriate replacement of $C^{(2)}$ is therefore
\be \label{eq:rpc2}
    C^{(2)} \rightarrow \sgn \bigl( p - q \, C^{(0)} \bigr) \, \bigl( \tilde{\Theta}^\intercal \, \tilde{\Sigma} \bigr)\,,
\ee
which is manifestly \SLZ-invariant and gives back $C^{(2)}$ when $\Theta = (1\,\,\,0)^\intercal$ and $\tilde{\Theta} = (0\,\,\,1)^\intercal$\,.

Applying the replacements \eqref{eq:rpc0}, \eqref{eq:rpphi}, \eqref{eq:rpf}, and \eqref{eq:rpc2} to  the D3-brane action \eqref{eq:dpactiond3}, we find the following manifestly \SLZ-invariant action:
\begin{empheq}[box=\fbox]{align} \label{eq:sd3bb}
\begin{split}
    S_{\text{D}3} = & - T_3 \int d^4 \sigma \, \sqrt{-\det
    \begin{pmatrix}
        0 &\,\,\,\, \tau^\text{\scalebox{0.8}{E}}_\beta \\[2pt]
        \bar{\tau}^\text{\scalebox{0.8}{E}}_\alpha &\,\,\,\, E^\text{\scalebox{0.8}{E}}_{\alpha\beta} + \frac{\Theta^\intercal \, \mathscr{F}_{\alpha\beta}}{\Theta^\intercal \, \CC}
    \end{pmatrix}} \\[2pt]
    & - T_3 \int \ls \CC^{(4)} + \bigl( \tilde{\Theta}^\intercal \, \tilde{\Sigma} \bigr) \wedge \bigl( \Theta^\intercal \, \mathscr{F} \bigr) - \frac{1}{2} \, \frac{\tilde{\Theta}^\intercal \, \CC}{\Theta^\intercal \, \CC} \, \bigl( \Theta^\intercal \, \mathscr{F} \bigr) \wedge \bigl( \Theta^\intercal \, \mathscr{F} \bigr) \rs.
\end{split}
\end{empheq}
We have introduced an \SLZ\, singlet $\CC^{(4)}$ that replaces $C^{(4)}$\,. We will later derive the explicit form of $\CC^{(4)}$ in Eq.~\eqref{eq:cc4c4} by analyzing the gauge symmetries. Unlike nonrelativistic $(p\,,q)$-strings, there is a unique nonrelativistic D3-brane action that is insensitive to the sign of $p- q \, C^{(0)}$\,. However, the SL($2\,, \mathbb{Z}$) transformations still have a branched structure as shown in Eqs.~\eqref{eq:tc} and \eqref{eq:spsgn}, \emph{i.e.}, the transformations of the two-form fields acquires a sign of the quantity $c \, C^{(0)} + d$\,. In this sense, the SL($2\,, \mathbb{Z}$) duality of D3-brane is also branched.

In order to compensate for the variation of the action \eqref{eq:sd3bb} with respect to the gauge transformation rule $\delta \tilde{\Sigma} = d\Xi$\,, the \SLZ-invariant $\CC^{(4)}$ has to transform under the Ramond-Ramond and Kalb-Ramond gauge symmetries as
\be \label{eq:gc4t}
    \delta \CC^{(4)} = dZ^{(3)} + \bigl( \Theta^\intercal \, d\Sigma \bigr) \wedge \bigl( \tilde{\Theta}^\intercal \, \Xi \bigr)\,,
\ee
where $Z^{(3)}$ is a parameter characterizing the three-form gauge transformation. The relation between $\CC^{(4)}$ and the objects that we have defined, namely, $C^{(4)}$ and $\Sigma$\,, can be fixed by matching the gauge transformations, which gives
\be \label{eq:cc4c4}
    \CC^{(4)} = C^{(4)} + \frac{1}{4} \,\Sigma^\intercal \wedge \bigl( \rho \, \Sigma \bigr) - \frac{1}{2} \, \bigl( \Theta^\intercal \, \Sigma \bigr) \wedge  \, \bigl( \tilde{\Theta}^\intercal \, \Sigma \bigr)\,.
\ee
Here, we defined the constant matrix
\begin{align}\label{eq:defUrho}
    \rho =
    \begin{pmatrix}
         0 & \,\,\, 1 \\[2pt]
        1 & \,\,\, 0
    \end{pmatrix}\,.
\end{align}
The gauge transformation of $C^{(4)}$ can be found in Eq.~\eqref{eq:rrgs}, where the gauge parameter $\zeta^{(3)}$ is related to $Z^{(3)}$ in Eq.~\eqref{eq:gc4t} via
\be \label{eq:Z3}
    Z^{(3)} = \zeta^{(3)} - \frac{1}{2} \, \Sigma^\intercal \wedge \bigl( \Theta \, \tilde{\Theta}^\intercal + \tilde{\Theta} \, \Theta^\intercal \bigr) \, \Xi + \frac{1}{2} \, \Sigma^\intercal \wedge \bigl( \rho \, \Xi \bigr)\,.
\ee
From the \SLZ-invariant expression \eqref{eq:cc4c4}, we can solve for the SL(2,\,$\mathbb{Z}$) transformation of $C^{(4)}$, which gives,
\be \label{eq:c4trnsf}
    C'{}^{(4)} = C^{(4)} + \frac{1}{2} \, \CK^\intercal
    \begin{pmatrix}
        b \, d &\,\,\, - b \, c \\[2pt]
        - b \, c &\,\,\, a \, c
    \end{pmatrix}
    \wedge \CK\,,
\ee
where
\be
    \CK = \begin{pmatrix}
        B \\[2pt]
        C^{(2)}
    \end{pmatrix} - \frac{c}{c \, C^{(0)} + d} \begin{pmatrix}
        0 \\[2pt]
        \frac{1}{2} \, \ell \, e^{-2\Phi}
    \end{pmatrix}\,.
\ee

{Note that the action \eqref{eq:sd3bb} becomes independent of $\tilde{\Theta}$ after the condition \eqref{eq:pqcond1} is imposed, up to a boundary term that we ignore.
Moreover, under the condition $\Theta = (1\,\,\,0)^\intercal$\,,} as expected, the action reduces to the original nonrelativistic D3-brane action \eqref{eq:dpactiond3}. Alternatively, when $\Theta = (0\,\,\,1)^\intercal$\,, the action \eqref{eq:sd3bb} becomes
\begin{align} \label{eq:tsd3}
\begin{split}
    \tilde{S}^{}_\text{D3} = & - T_3 \int d^4 \sigma \, e^{-\Phi} \sqrt{- \det
    \begin{pmatrix}
        0 &\,\,\,\, \tau_\beta
 \\[2pt]
 \bar{\tau}_\alpha &\,\,\,\, E_{\alpha\beta} + \tilde{\CF}_{\alpha\beta}
    \end{pmatrix}} \\[2pt]
    & - T_3 \int \lr \tilde{C}^{(4)} + \tilde{C}^{(2)} \wedge \tilde{\CF} + \frac{1}{2} \, \tilde{C}^{(0)} \, \tilde{\CF} \wedge \tilde{\CF} \rr,
\end{split}
\end{align}
where
\begin{subequations}
\begin{align}
    \tilde{C}^{(0)} & = - C^{(0)},
        &%
    \tilde{\CF} & = - \frac{C^{(2)}+dA^{\text{\scalebox{0.8}{C}}}}{C^{(0)}} + \frac{\ell}{2 \, \bigl( e^{\Phi} \, C^{(0)} \bigr)^2}\,, \\[4pt]
    \tilde{C}^{(2)} & = C^{(0)} \ls B - \frac{\ell}{2 \, \bigl( e^{\Phi} \, C^{(0)} \bigr)^2} \rs,
        &%
    \tilde{C}^{(4)} & = C^{(4)} + \lr C^{(2)} - \frac{e^{-2\Phi} \, \ell}{2 \, C^{(0)}} \rr \wedge B\,.
\end{align}
\end{subequations}
This is the S-dual action of the original nonrelativistic D3-brane action \eqref{eq:dpactiond3}, which can be equivalently obtained by
dualizing the $U(1)$ gauge field on the D3-brane \cite{Ebert:2021mfu}.

\subsection{Inter-Branch D3-Branes} \label{sec:ibd3b}

In Section~\ref{sec:nrd1b}, we found the actions $S^\pm$ in Eqs.~\eqref{eq:spq} and \eqref{eq:spqm}, which describe nonrelativistic $(p\,, q)$-strings satisfying $p - q \, C^{(0)} \neq 0$\,. 
Although the $(p\,,q)$-string action is well defined in both the branches, it becomes singular at the boundary defined by $p - q \, C^{(0)} = 0$\,.
In contrast, intriguingly, when the D3-brane action \eqref{eq:sd3bb} is concerned, the limit $p - q \, C^{(0)} \rightarrow 0$ appears to be non-singular (\emph{e.g.}, see \cite{Ebert:2021mfu} when $\Theta = (0\,\,\,1)^\intercal$).
In the following, we study explicitly such inter-branch D3-branes and uncover the resulting finite action in the $p - q \, C^{(0)} \rightarrow 0$ limit.

We start with defining $\delta \equiv p - q \, C^{(0)}$ and thus consider the $\delta \rightarrow 0$ limit. Expanding Eq.~\eqref{eq:sd3bb} with respect to $\delta$\,, we find that the DBI-like part of the Lagrangian gives
\begin{align}
\begin{split}
	\CL_\text{DBI} \equiv & - \sqrt{-\det
	\begin{pmatrix}
		0 &\,\,\, \tau^\text{\scalebox{0.8}{E}}_\beta \\[2pt]
		\bar{\tau}^\text{\scalebox{0.8}{E}}_\alpha &\,\,\, E^\text{\scalebox{0.8}{E}}_{\alpha\beta} + \frac{\Theta^\intercal \, \mathscr{F}_{\alpha\beta}}{\Theta^\intercal \, \CC}
	\end{pmatrix}} \\[4pt]
	= &
	- \frac{1}{2 \, \delta^2} \frac{\bigl| q \, \tr \bigl( \star \ell \, \mathcal{X} \bigr) \bigr|}{e^{2\Phi}}
	- \frac{1}{4 \, \delta \, q} \, \tr \bigl( \star \mathcal{X} \, \mathcal{X} \bigr) \, \sgn \bigl[ q \, \tr \bigl( \star \ell \, \mathcal{X} \bigr) \bigr] \\[4pt]
	& - \frac{q^2 \, e^{-2\Phi} \, G + \tr \bigl( \, \star \mathcal{X} \, \tau \star \! \mathcal{X} \, E \, \bigr) - \frac{1}{16} \, q^{-2} \, e^{2\Phi} \, \Bigl[ \tr \bigl( \star \mathcal{X} \, \mathcal{X} \bigr) \Bigr]^2}{\bigl| q \, \tr \bigl( \star \ell \, \mathcal{X} \bigr) \bigr|} + O (\delta)\,.
\end{split}
\end{align}
Since the worldvolume is flat, the indices are raised (lowered) by the Minkowski metric $
\eta^{\alpha\beta}$ ($\eta_{\alpha\beta}$).
Here, $( \tau )_{\alpha\beta} = \tau_{\alpha\beta}$\,,
$( E )_{\alpha\beta} = E_{\alpha\beta}$\,, and
\begin{subequations}
    \begin{align}
    	\mathcal{X} &=
	p \, \bigl( B + dA^{\text{\scalebox{0.8}{B}}} \bigr) +
	q \, \bigl( C^{(2)}+dA^{\text{\scalebox{0.8}{C}}} \bigr)\,,\\[2pt]
    G &= - \frac{1}{4} \, \epsilon^{\alpha_1 \alpha_2 \alpha_3 \alpha_4} \, \epsilon^{\beta_1 \beta_2 \beta_3 \beta_4} \, \tau^{}_{\alpha_1\beta_1} \, \tau^{}_{\alpha_2\beta_2} \, E^{}_{\alpha_3\beta_3} \, E^{}_{\alpha_4\beta_4}\,.
\end{align}
\end{subequations}
The Hodge duals of $\ell$ and $\mathcal{X}$ are defined to be
\be
    \bigl( \star \ell \bigr)_{\alpha\beta} = \tfrac{1}{2} \, \epsilon_{\alpha\beta}{}^{\gamma\delta} \, \ell_{\gamma\delta}\,,
        \qquad%
    \bigl( \star \mathcal{X} \bigr)_{\alpha\beta} = \tfrac{1}{2} \, \epsilon_{\alpha\beta}{}^{\gamma\delta} \, \mathcal{X}_{\gamma\delta}\,.
\ee
Similarly, the Wess-Zumino part of the Lagrangian in Eq.~\eqref{eq:sd3bb} gives
\begin{align}
\begin{split}
	\mathcal{L}_\text{WZ} \equiv & - \ls \CC^{(4)} + \bigl( \tilde{\Theta}^\intercal \, \tilde{\Sigma} \bigr) \wedge \bigl( \Theta^\intercal \, \mathscr{F} \bigr) - \frac{1}{2} \, \frac{\tilde{\Theta}^\intercal \, \CC}{\Theta^\intercal \, \CC} \, \bigl( \Theta^\intercal \, \mathscr{F} \bigr) \wedge \bigl( \Theta^\intercal \, \mathscr{F} \bigr) \rs \\[4pt]
	= & - \frac{1}{\delta^2} \frac{q}{e^{2\Phi}} \, \ell \wedge \mathcal{X} - \frac{1}{2 \, \delta \, q} \, \mathcal{X} \wedge \mathcal{X} \\[2pt]
	& - \mathcal{X}^{(4)} - \bigl( \, \tilde{p} \, B + \tilde{q} \, C^{(2)} \bigr) \wedge \mathcal{X} + \frac{\tilde{q}}{2 \, q} \, \mathcal{X} \wedge \mathcal{X} + O(\delta).
\end{split}
\end{align}
Here,
\be
    \mathcal{X}^{(4)} = C^{(4)} + \frac{1}{2} \, B \wedge C^{(2)} - \frac{1}{2} \, \bigl( p \, B + q \, C^{(2)} \bigr) \wedge \bigl( \tilde{p} \, B + \tilde{q} \, C^{(2)} \bigr)\,,
\ee
in analogy with the definition \eqref{eq:cc4c4}.
For the combination $\CL_\text{DBI} + \CL_\text{WZ}$ to be finite, we require
\be \label{eq:qlx}
	q \, \tr \bigl( \star \ell \, \mathcal{X} \bigr) > 0\,.
\ee
In particular, we require that $q \neq 0$\,. As a result, in the limit $\delta \rightarrow 0$\,, we find a nonsingular action for the inter-branch D3-brane as
\begin{align} \label{eq:ibb3a}
\begin{split}
	S = & \,\, - T_3 \int d^4 \sigma \, \frac{q^2 \, e^{-2\Phi} \, G + \tr \bigl( \, \star \mathcal{X} \, \tau \star \! \mathcal{X} \, E \, \bigr) - \frac{1}{16} \, q^{-2} \, e^{2\Phi} \, \Bigl[ \tr \bigl( \star \mathcal{X} \, \mathcal{X} \bigr) \Bigr]^2}{q \, \tr \bigl( \star \ell \, \mathcal{X} \bigr)} \\[4pt]
	& - T_3 \int \ls \mathcal{X}^{(4)} + \bigl( \, \tilde{p} \, B + \tilde{q} \, C^{(2)} \bigr) \wedge \mathcal{X} - \frac{\tilde{q}}{2 \, q} \, \mathcal{X} \wedge \mathcal{X} \rs.
\end{split}
\end{align}
Here, the quantity $|q \, \tr \bigl( \star \ell \, \mathcal{X}) |$ controls the size of the effective coupling of the inter-branch D3-brane. By construction, the action \eqref{eq:ibb3a} has to be invariant under the branced SL(2,\,$\mathbb{Z}$) transformations in Eqs.~\eqref{eq:slztfs} and \eqref{eq:c4trnsf}. In particular, note that $e^{-\Phi} \, q \, \tr \bigl( \star \ell \, \mathcal{X})$ is SL(2,\,$\mathbb{Z}$) invariant, which implies that the inequality \eqref{eq:qlx} is preserved under the branched SL(2,\,$\mathbb{Z}$) transformations.

In the case where $\Theta = (0,\,1)$\,, the action \eqref{eq:ibb3a} becomes
\begin{align} \label{eq:sncym}
\begin{split}
	S = & \,\, - T_3 \int d^4 \sigma \, \frac{e^{-2\Phi} \, G + \tr \bigl( \, \star \mathcal{X} \, \tau \star \! \mathcal{X} \, E \, \bigr) - \frac{1}{16} \, e^{2\Phi} \, \Bigl[ \tr \bigl( \star \mathcal{X} \, \mathcal{X} \bigr) \Bigr]^2}{\tr \bigl( \star \ell \, \mathcal{X} \bigr)} \\[2pt]	%
	& - T_3 \int \ls C^{(4)} + B \wedge \bigl( C^{(2)} + d A^{\text{\scalebox{0.8}{C}}} \bigr) \rs.
\end{split}
\end{align}
This is the action that has been studied in \cite{Ebert:2021mfu}, which is S-dual to the nonrelativistic D3-brane action \eqref{eq:dpactiond3} at $C^{(0)} \rightarrow 0$\,. In other words, Eq.~\eqref{eq:sncym} coincides with the S-dual action \eqref{eq:tsd3} in the $C^{(0)} \rightarrow 0$ limit.
This theory has a sector that describes noncommutative Yang-Mills (NCYM) theory \cite{Gopakumar:2000na, Ebert:2021mfu}. In order to make this NCYM sector manifest, we focus on flat spacetime and choose the following background field configuration:
\be
    \tau_\mu{}^A = \delta_\mu^A\,,
    \qquad
    E_\mu{}^{A'} = \delta_\mu^{A'}\,,
    \qquad C^{(q)}=0\,,
    \qquad
    B_{\alpha\beta} = -
    \begin{pmatrix}
        \mathbb{0}_{2\times2} &\,\, \mathbb{0}_{2\times2} \\[2pt]
        \mathbb{0}_{2\times2} &\,\, b \, \epsilon^{}_{ij}
    \end{pmatrix}\,,
\ee
where $i\,,j=2\,,3$. Here, $b$ is a constant and  $\epsilon^{}_{ij}$ is a Levi-Civita symbol defined by $\epsilon^{}_{23} = - \epsilon^{}_{32} = 1$\,. Moreover, we choose the D-brane configuration such that $X^\alpha = \sigma^\alpha$\,, $\alpha = 0, \cdots, 3$\,, such that the D3-brane extends in the target-space longitudinal directions. Using the Seiberg-Witten map \cite{Seiberg:1999vs}, we find that the effective open string coupling is given by
\be \label{eq:Go}
    G_\text{o} = \sqrt{2\pi g_s \, b}\,,
\ee
where $g_s = e^{\langle \Phi \rangle}$ is the closed string coupling. Moreover, the worldvolume acquires noncommutativity in the $\sigma^2$ and $\sigma^3$ directions, with
\be
    [\sigma^2\,, \sigma^3] \propto \frac{1}{b}\,,
\ee
controlled by the inverse of the ``magnetic" $B$-field. This implies that the resulting theory in the chosen configuration describes NCYM on the worldvolume, which is S-dual to noncommutative open string theory  \cite{Gopakumar:2000na}.\,\footnote{The fact that the S-dual relation in \cite{Gopakumar:2000na} between NCYM and noncommutative open string (NCOS) theory only arises when the inter-branch condition $p - q \, C^{(0)} = 0$ is satisfied has been observed in \emph{e.g.} \cite{Gran:2001tk}.} The NCYM coupling is given by the open string coupling $G_\text{o}$ in Eq.~\eqref{eq:Go}, which is proportional to the square-root of the magnetic $B$-field.


\section{Conclusions} \label{sec:concl}

In this paper, we constructed manifestly \SLZ-invariant actions in nonrelativistic string theory for D-instantons in Eq.~\eqref{eq:sld-1}, $(p\,, q)$-strings in Eqs.~\eqref{eq:spq} and \eqref{eq:spqm} and D3-branes in Eq.~\eqref{eq:sd3bb}, where the branes are coupled to a non-Lorentzian background geometry equipped with a codimension-two foliation structure. Intriguingly, for nonrelativistic $(p\,,q)$-strings, such a realization requires a branching in the $(p\,,q)$-space, determined by the sign of $p - q \, C^{(0)}$, with $C^{(0)}$ the zero-form axion background field. The states in the different branches are described by distinct actions, related to each other by sending $(p\,, q)$ to $(-p\,, -q)$\, without changing the background fields.
The SL(2,\,$\mathbb{Z}$) group itself is also branched. Consider the SL(2,\,$\mathbb{Z}$) group element $\Lambda$ in Eq.~\eqref{eq:sl2zpar},
which satisfies $c \, C^{(0)} + d > 0$\,. Such a transformation maps covariantly each of the branches in $(p\,,q)$-space to itself. In contrast, an SL(2,\,$\mathbb{Z}$) transformation satisfying $c \, C^{(0)} + d < 0$ maps between one branch in $(p\,,q)$-space and the other.

We also discussed the inter-branch D3-branes that arise from the limit $p - q \, C^{(0)} \rightarrow 0$\,, resulting in the non-singular, \SLZ-invariant brane action \eqref{eq:ibb3a}. In flat spacetime with zero Ramond-Ramond potentials and $(p\,,q)=(0\,,1)$\,, the inter-branch D3-brane action leads to noncommutative Yang-Mills theory \cite{Gopakumar:2000na, Ebert:2021mfu}.

It should be noted that the non-relativistic SL(2,\,$\mathbb{Z}$) transformations we constructed in Eqs.~\eqref{eq:slztfs} and \eqref{eq:c4trnsf} are a deformation of the standard relativistic duality rules containing extra terms. It is well-known that these relativistic SL(2,\,$\mathbb{Z}$) duality transformations have a geometric interpretation, from a nine-dimensional point of view, as the geometric transformations of an M-theory two-torus \cite{Schwarz:1995dk}. It would be interesting to see what the geometric interpretation of the extra terms in the non-relativistic SL(2,\,$\mathbb{Z}$) duality rules are, given the fact that the general coordinate transformations do not have non-relativistic corrections.

The fact that the SL(2,\,$\mathbb{Z}$) duality in nonrelativistic string theory involves a branching depending on $C^{(0)}$ may suggest that a background D7-brane is needed for a thorough understanding of this branched duality. The magnetic dual of $\Cz$ is an eight-form field, which naturally couples to the D7-brane. It would therefore be natural to examine the role that such D7-branes might play in describing the dynamics of $(p\,,q)$-strings in nonrelativistic string/F-theory.
Moreover, it would be interesting to study the implications of our studies of various bound states for the Hagedorn transition along the lines of \cite{Gubser:2000mf}.

It is important to extend the analysis of this project to include supersymmetry and study the corresponding type IIB supergravity background fields. This will be an extension of the construction of nonrelativistic minimal supergravity in \cite{Bergshoeff:2021bmc,Bergshoeff:2021tfn} to the case of maximal supersymmetry. It would be intriguing to understand the realization of the branched SL(2,\,$\mathbb{Z}$) symmetry in that theory.
Once the SL(2,\,$\mathbb{Z}$) duality is understood in supergravity, it is natural to apply it together with the nonrelativistic T-duality transformations \cite{Bergshoeff:2018yvt, Ebert:2021mfu} to construct new half-BPS background solutions out of known solutions, such as a nonrelativistic version of the well-known D-brane solutions, along the lines of \cite{Bergshoeff:2022pzk}. We expect that these solutions will provide an important input for a top-down construction of nonrelativistic holography. We are currently working on a partner article addressing these questions.

\acknowledgments

We would like to thank Jan Rosseel, Ceyda \c{S}im\c{s}ek, and Patrik Novosad for discussions on S-duality. Z.Y. would like to thank Stephen Ebert and Hao-Yu Sun for useful discussions.
K.T.G. has received funding from the European Union’s Horizon 2020 research and innovation programme under the Marie Sk\l odowska-Curie grant agreement No 101024967. Z.Y. is supported by the European Union’s Horizon 2020 research and innovation programme under the Marie Sk{\l}odowska-Curie grant agreement No 31003710. U.Z. is supported by TUBITAK - 2218 National Postdoctoral Research
Fellowship Program with grant number 118C512. Nordita is supported in part by NordForsk.

\newpage

\appendix

\section{Derivation of \texorpdfstring{SL(2,\,$\mathbb{Z}$)}{SLZ} Invariants} \label{app:ivs}

In our construction of the manifestly \SLZ-invariant nonrelativistic D3-brane action in Section~\ref{sec:nd3b}, the following \SLZ-invariant is introduced in Eq.~\eqref{eq:rpc2}:
\be \label{eq:spqc0ts}
    \sgn \bigl( p - q \, C^{(0)} \bigr) \, \bigl( \tilde{\Theta}^\intercal \, \tilde{\Sigma} \bigr)\,,
\ee
where
$\tilde{\Theta} = \bigl( \,\tilde{p}\,\,\,\, \tilde{q} \, \bigr)^\intercal$ and
\be \label{eq:tss}
    \tilde{\Sigma} =
    \Sigma
    +
    \frac{q \, \tilde{q}{\,}^{-1}}{\bigl( p - q \, C^{(0)} \bigr)^2}
    \begin{pmatrix}
        0 \\[2pt]
        \frac{1}{2} \, \ell \, e^{-\Phi}
    \end{pmatrix},
        \qquad%
    \Sigma =
    \begin{pmatrix}
        B \\[2pt]
        C^{(2)}
    \end{pmatrix} +
    \frac{q}{p - q \, C^{(0)}}
    \begin{pmatrix}
        0 \\[2pt]
        \frac{1}{2} \, \ell \, e^{-2\Phi}
    \end{pmatrix}.
\ee
It is straightforward to check the invariance of Eq.~\eqref{eq:spqc0ts} under the SL(2,\,$\mathbb{Z}$) transformations \eqref{eq:slztfs}. In this appendix, we present a systematic procedure of constructing the \SLZ-invariants that appeared in the bulk of the paper by starting with the transformation rules of the background fields, using which the invariant \eqref{eq:spqc0ts} can also be naturally derived.

\subsection{Branched \texorpdfstring{SL(2,\,$\mathbb{Z}$) }{SLZ} Transformations}

We start with summarizing the SL(2,\,$\mathbb{Z}$) transformations
that have been given in Eqs.~\eqref{eq:slztfs} and \eqref{eq:c4trnsf}, which act on the Ramond-Ramond potentials and the $B$-field as follows:\,\footnote{It is interesting to note the curious fact that the transformation \eqref{eq:cp4app} of $C^{(4)}$ is reminiscent of its counterpart in relativistic string theory \cite{Meessen:1998qm}.
However, $\CK$ is \emph{not} an \SLZ-doublet: $\CK$ is only defined with respect to a particular SL(2,\,$\mathbb{Z}$) transformation, with explicit dependence on the Lie group parameters $c$ and $d$\,.
Moreover, the transformation \eqref{eq:slztbc} of the two-form fields can be brought into the form analogous to the relativistic case as
\be
    \CK' = \sgn (c \, C^{(0)} + d) \, \bigl(\Lambda^{-1}\bigr)^\intercal \, \CK\,.
\ee
Note that $\CK'$ contains $c' = - c$ and $d' = a$\,.
Here, $\CK'$ depends on the SL(2,\,$\mathbb{Z}$) transformation $\Lambda'$ that maps the primed fields to the unprimed fields, \emph{i.e.}, $\Lambda' = \Lambda^{-1}$\,.}
\begin{subequations} \label{nrsltrans}
\begin{align}
    C'{}^{(0)} & = \frac{a \, C^{(0)} + b}{c \, C^{(0)} + d}\,, \label{eq:lc0phi} \\[2pt]
    \begin{pmatrix}
        B' \\[2pt]
        C'{}^{(2)}
    \end{pmatrix}
    & = \text{sgn} \bigl( c \, C^{(0)} + d \bigr) \,
    \ls \bigl( \Lambda^{-1} \bigr)^\intercal \, \CK - \frac{c}{\bigl( c \, C^{(0)} + d \bigr)^2}
    \begin{pmatrix}
        0 \\[2pt]
        \frac{1}{2} \, \ell \, e^{-2\Phi}
    \end{pmatrix} \rs, \label{eq:slztbc} \\[2pt]
    C'{}^{(4)} & = C^{(4)} + \frac{1}{2} \, \CK^\intercal
    \begin{pmatrix}
        b \, d &\,\,\, - b \, c \\[2pt]
        - b \, c &\,\,\, a \, c
    \end{pmatrix}
    \wedge \CK\,, \label{eq:cp4app}
\end{align}
\end{subequations}
where
\be \label{eq:defk}
    \CK = \begin{pmatrix}
        B \\[2pt]
        C^{(2)}
    \end{pmatrix} - \frac{c}{c \, C^{(0)} + d} \begin{pmatrix}
        0 \\[2pt]
        \frac{1}{2} \, \ell \, e^{-2\Phi}
    \end{pmatrix}\,.
\ee
It is also useful to recall that
\be
    \ell' = |c \, C^{(0)} + d| \, \ell\,,
        \qquad%
    \Phi' = \Phi + 2 \, \ln |c \, C^{(0)} + d|\,, \label{nrsl3}
\ee
which are needed for explicitly checking the \SLZ-invariance of the quantities that we will construct in the following.

\subsection{Zero-Form Invariants} \label{sec:zfi}

In order to demonstrate the procedure of deriving \SLZ-invariants, we start with using the SL(2,\,$\mathbb{Z}$) transformation of the zero-form $C^{(0)}$ in Eq.~\eqref{eq:lc0phi} to derive a family of SL(2,\,$\mathbb{Z}$) invariants. Recall that $\Theta$ and $\tilde{\Theta}$ are both \SLZ-doublets, which transform under the action of the \SLZ \, matrix $\Lambda$ as
\be
    \Theta' = \Lambda \, \Theta\,,
        \qquad%
    \tilde{\Theta}' = \Lambda \, \tilde{\Theta}\,.
\ee
First, we fix all the primed background fields and the primed doublets $\Theta'$ and $\tilde{\Theta}'$\,, which take the following reference values:
\be \label{eq:tttxy}
    \Theta' = \begin{pmatrix}
        x \\[2pt]
        y
    \end{pmatrix}\,,
        \qquad%
    \tilde{\Theta}' =
    \begin{pmatrix}
        \tilde{x} \\[2pt]
        \tilde{y}
    \end{pmatrix}\,,
\ee
where the constant integers $x,\,  y, \, \tilde{x}, \, \tilde{y}$ satisfy $x \, \tilde{y} - y \, \tilde{x} = 1$\,.
Then, the matrix $\Lambda$ can be expressed in terms of $\Theta$ and $\tilde{\Theta}$\,, with
\be \label{eq:lpq}
    \Lambda =
    \begin{pmatrix}
        x \, \tilde{q} - \tilde{x} \, q &\,\,\, \tilde{x} \, p - x \, \tilde{p} \, \\[2pt]
        y \, \tilde{q} - \tilde{y} \, q &\,\,\, \tilde{y} \, p - y \, \tilde{p} \,
    \end{pmatrix}\,.
\ee
Plugging Eq.~\eqref{eq:lpq} into the transformation rule \eqref{eq:lc0phi} for $C^{(0)}$\,, we find
\be \label{eq:cp0inv}
    C'{}^{(0)} =\frac{\tilde{x} \, \bigl( p - q \, C^{(0)} \bigr) - x \, \bigl( \tilde{p} - \tilde{q} \, C^{(0)} \bigr)}{y \, \bigl( \tilde{p} - \tilde{q} \, C^{(0)} \bigr) - \tilde{y} \, \bigl( p - q \, C^{(0)} \bigr)}\,.
\ee
Since we have fixed the primed background fields including the zero-form $C'{}^{(0)}$\,, the RHS of Eq.~\eqref{eq:cp0inv} takes the same value for any $\Theta$ and $\tilde{\Theta}$\,, which are related to the fixed $\Theta'$ and $\tilde{\Theta}'$ in Eq.~\eqref{eq:tttxy} via an SL(2,\,$\mathbb{Z}$) transformation. Therefore, due to the associativity of group actions,
the RHS of Eq.~\eqref{eq:cp0inv} defines an SL(2,\,$\mathbb{Z}$) invariant.

Note that the invariant implied by Eq.~\eqref{eq:cp0inv} is defined with respect to the arbitrary but fixed reference parameters in Eq.~\eqref{eq:tttxy}, which ensures that it is \SLZ-invariant, but does not necessarily mean that it must arise in the D-instanton Lagrangian in nonrelativistic string theory. The actual terms that arise in the nonrelativistic D$(-1)$-brane action correspond to fixing the reference values of $\Theta'$ and $\tilde{\Theta}'$ in Eq.~\eqref{eq:tttxy} to be
\be \label{eq:0110}
    \Theta' =
    \begin{pmatrix}
        1 \\[2pt]
        0
    \end{pmatrix}\,,
        \qquad%
    \tilde{\Theta}' =
    \begin{pmatrix}
        0 \\[2pt]
        1
    \end{pmatrix}\,.
\ee
Then, the SL(2,\,$\mathbb{Z}$) matrix Eq.~\eqref{eq:lpq} becomes
\be \label{eq:lsimpq}
    \Lambda =
    \begin{pmatrix}
        \tilde{q} &\,\, - \tilde{p} \, \\[2pt]
        - q &\,\,\, p
    \end{pmatrix}\,.
\ee
Henceforth, we will fix $\Theta'$ and $\tilde{\Theta}'$ to the values in Eq.~\eqref{eq:0110}.
In this case, the \SLZ-invariant implied by Eq.~\eqref{eq:cp0inv} becomes
\be
    \frac{\tilde{p} - \tilde{q} \, C^{(0)}}{p - q \, C^{(0)}}\,,
\ee
which reproduces the \SLZ-invariant that we already constructed in Eq.~\eqref{eq:sld-1} for nonrelativistic D-instantons.

\subsection{Higher-Form Invariants} \label{sec:tfi}

We are now ready to apply the same method developed in Section~\ref{sec:zfi} to the \SLZ \, transformations of the higher-form fields and derive the associated SL(2,\,$\mathbb{Z}$) invariants. Again, the actual terms that arise in the D$p$-brane action correspond to fixing $\Theta'$ and $\tilde{\Theta}'$ as in Eq.~\eqref{eq:0110}.

We first 
consider the \SLZ \, transformations of the two-form fields in Eq.~\eqref{eq:slztbc}. Using the reference values of $\Theta'$ and $\tilde{\Theta}'$ in Eq.~\eqref{eq:0110}, and plugging Eq.~\eqref{eq:lsimpq} into Eq.~\eqref{eq:slztbc}, we find
\begin{align}
    \begin{pmatrix}
        B' \\[2pt]
        C'{}^{(2)}
    \end{pmatrix}
    =
    \sgn \bigl(p - q \, C^{(0)} \bigr)
    \begin{pmatrix}
        \Theta^\intercal \, \Sigma \\[2pt]
        \tilde{\Theta}^\intercal \, \tilde{\Sigma}
    \end{pmatrix}\,,
\end{align}
where $\Sigma$ and $\tilde{\Sigma}$ are given in Eq.~\eqref{eq:tss}. Since the primed background fields $B'$ and $C'{}^{(2)}$ are both fixed, we find two independent \SLZ-invariants,
\be \label{eq:tfinv}
    \sgn \bigl( p - q \, C^{(0)} \bigr) \, \bigl( \Theta^\intercal \, \Sigma \bigr)\,,
        \qquad%
    \sgn \bigl( p - q \, C^{(0)} \bigr) \, \bigl( \tilde{\Theta}^\intercal \, \tilde{\Sigma} \bigr)\,.
\ee
The first expression in Eq.~\eqref{eq:tfinv} reproduces the Wess-Zumino term in the nonrelativistic $(p\,, q)$-string action \eqref{eq:ta}. Moreover, the second expression in Eq.~\eqref{eq:tfinv} gives the \SLZ-invariant that we quoted in Eq.~\eqref{eq:rpc2}, which is important to the construction of the nonrelativistic D3-brane action \eqref{eq:sd3bb}.

Finally, plugging Eq.~\eqref{eq:lsimpq} into the SL(2,\,$\mathbb{Z}$) transformation \eqref{eq:cp4app} of the four-form field $C^{(4)}$\,, we find
\be
    C'{}^{(4)} = C^{(4)} - \frac{1}{2} \, \Sigma^\intercal
    \begin{pmatrix}
        p \, \tilde{p} &\,\,\, q \, \tilde{p} \, \\[2pt]
        q \, \tilde{p} &\,\,\, q \, \tilde{q} \,
    \end{pmatrix} \wedge \Sigma = \CC^{(4)}\,,
\ee
where $\CC^{(4)}$ is defined in Eq.~\eqref{eq:cc4c4}, which is the \SLZ-singlet desired for the construction of the nonrelativistic D3-brane action \eqref{eq:sd3bb}.

\section{Nonrelativistic String Limit}\label{sec:NRlimits}

In this appendix, we show how the D-instanton action \eqref{eq:sld-1}, the $(p,\,q)$-string actions \eqref{eq:spq} and \eqref{eq:spqm}, and the D3-brane action \eqref{eq:sd3bb} in nonrelativistic string theory follow from the nonrelativistic string limit of relativistic string theory. We start by reviewing the corresponding actions for relativistic instantons, $(p,\,q)$-strings, and D3-branes, which are given by \cite{Townsend:1997kr, Bergshoeff:2006gs}
\begin{subequations}\label{eq:relbranes}
    \begin{align}
    \hat{S}_\text{D(-1)} &= T_{-1} \, \frac{\tilde{\Theta}^\intercal \, \hat{\CM}^{-1} \, \Theta}{\Theta^\intercal \, \hat{\CM}^{-1} \, \Theta}\,,\label{eq:relinst}\\[.2truecm]
    \hat S_\text{string} &= - T^{}_1 \int d^2 \sigma \, \sqrt{\Theta^\intercal \, \hat{\CM}^{-1} \, \Theta}\, \sqrt{- \det \hat{G}^{\text{\scalebox{0.8}{E}}}_{\alpha\beta}} \, - T^{}_1 \int \Theta^\intercal \, \hS\,,\label{eq:relpqstring}\\[6pt]
    \hat{S}_{\text{D}3} & = - \, T_3 \int d^4 \sigma \, \sqrt{-\det \! \lr {\hat{G}}^{\text{\scalebox{0.8}{E}}} + \frac{\Theta^\intercal \, \hat{\CF}}{\sqrt{\Theta^\intercal \, \hat{\CM}^{-1} \, \Theta}} \rr} \notag \\[2pt]
    & \quad - T_3 \int \ls \hat{\CC}^{(4)} + \bigl( \tilde{\Theta}^\intercal \, \hat{\Sigma} \bigr) \wedge \bigl( \Theta^\intercal \, \hat{\CF} \bigr) - \frac{1}{2} \, \frac{\tilde{\Theta}^\intercal \,\hat{\CM}^{-1} \, \Theta}{\Theta^\intercal \, \hat{\CM}^{-1} \, \Theta} \, \bigl( \Theta^\intercal \, \hat{\CF} \bigr) \wedge \bigl( \Theta^\intercal \, \hat{\CF} \bigr) \rs,\label{eq:relD3}
\end{align}
\end{subequations}
The \SLZ\, doublets $\Theta$ and $\tilde \Theta$ have already been defined in the bulk of the paper, see equations \eqref{eq:tcs}, \eqref{eq:tth}, and \eqref{eq:pqcond1}. The Einstein frame metric and its pullback are defined as
\be
    \hat{G}^\text{\scalebox{0.8}{E}}_{\alpha\beta} = \p_\alpha X^\mu \, \p_\beta X^\nu \, \hat{G}^\text{\scalebox{0.8}{E}}_{\mu\nu}\,,
        \qquad%
    \hat{G}^\text{\scalebox{0.8}{E}}_{\mu\nu} = e^{-\hat{\Phi}/2} \, \hat{G}_{\mu\nu}\,,
\ee
where $\alpha, \beta$ are the respective worldvolume indices. We furthermore use the following definitions of the matrix $\hat{\mathcal M}$ and the doublet of two-forms $\hat{\Sigma}$\,:
\be \label{eq:lgts}
    \hat{\CM} = e^{\hat{\Phi}}
    \begin{pmatrix}
        \bigl( \hat{C}^{(0)} \bigr)^2 + e^{-2 \hat{\Phi}} &\,\, \hat{C}^{(0)} \\[4pt]
        \hat{C}^{(0)} &\,\, 1
    \end{pmatrix}\,,
        \qquad%
    \hat\Sigma =
        \begin{pmatrix}
            \hat B\\[2pt]
            \hat C^{(2)}
        \end{pmatrix}\,.
\ee
Together with a doublet of Born-Infeld vectors $\mathcal A$, we also introduced the following worldvolume two-form $\hat\CF$ and \SLZ\, singlet four-form 
$\hat{\CC}^{(4)}$\,:
\begin{align}
    \hat{\CF} =
    \hat{\Sigma} + d\CA\,,
        \qquad%
    \hat{\CC}^{(4)} = \hat{C}^{(4)} + \frac{1}{2} \, \hat{B}\wedge\hat C^{(2)} - \frac{1}{2} \, \bigl( \Theta^\intercal \, \hat{\Sigma} \bigr) \wedge \bigl( \tilde{\Theta}^\intercal \, \hat{\Sigma} \bigr)\,.
\end{align}
The instanton, $(p,\,q)$ string, and D3 brane actions \eqref{eq:relbranes} are invariant\footnote{In the literature, the action \eqref{eq:relpqstring} is sometimes called covariant \cite{Townsend:1997kr}, when the action of \SLZ\,on $\Theta$ is considered to be passive.} under the action of \SLZ, where $\Theta$ and $\tilde\Theta$ transform as doublets, while the Einstein frame metric $\hat{G}^\text{\scalebox{0.8}{E}}$ and the four-form potential $\hat{\CC}{}^{(4)}$ transform as singlets. Moreover, the quantities $\hat\CM$\,, $\hat\Sigma$\,, and $\hat\CF$ transform as
\begin{align} \label{sltran}
    &\hat{\CM}' = \Lambda \, \hat{\CM} \, \Lambda^\intercal\,, && \hat{\Sigma}' = \big(\Lambda^{-1}\big)^\intercal\hat \Sigma\,,&& \hat\CF' = \big(\Lambda^{-1}\big)^\intercal\hat\CF\,.
\end{align}
In terms of the string coupling $\hat{g}_s = \langle e^{\hat{\Phi}} \rangle$\,, the tension of the $(p,\,q)$-string action \eqref{eq:relpqstring} is given by
\be \label{eq:htpq}
    \hat{T}_{p,\,q} = T_1 \sqrt{\Bigl( p - q \, \bigl\langle \hat{C}^{(0)} \bigr\rangle \Bigr)^2 + \frac{q^2}{\hat{g}_s^2}}\,,
\ee
which matches the result from type IIB supergravity \cite{Schwarz:1995dk}.

Nonrelativistic string theory arises from the nonrelativistic limit of relativistic string theory \cite{Gomis:2000bd}. To facilitate such a limit, we introduce a real dimensionless parameter $\omega>0$ by reparametrizing the relativistic metric field $\hat{G}_{\mu\nu}$\,, Kalb-Ramond field $\hat{B}_{\mu\nu}$\,, dilaton field $\hat{\Phi}$, and Ramond-Ramond forms $\hat{C}^{(q)}$ as follows \cite{Harmark:2019upf, Bergshoeff:2019pij, Ebert:2021mfu}:\,\footnote{It is understood that $C^{(q)} = 0$ for $q < 0$\,. In particular, for the purpose of this paper, $C^{(-2)} = 0$\,.}
\begin{subequations} \label{eq:TSrescal}
\begin{align}
    \hat G_{\mu\nu} &= \omega^2\,\tau_{\mu\nu} + E_{\mu\nu}\,,
        \qquad%
    \hat{\Phi} = \Phi + \ln \omega \,,\\[2pt]
    \hat{B} &= - \omega^2\,\ell + B\,, \label{eq:Bhatmunu} \\[2pt]
    \hat{C}^{(q)} &= \omega^2 \, C^{(q-2)} \wedge \ell + C^{(q)}\,.
\end{align}
\end{subequations}
Here, we define the two-form $\ell$ to be in the target space as opposed to Eq.~\eqref{eq:TSrescalC}, which is on the worldsheet:
\be
    \ell_{\mu\nu} = \tau_\mu{}^A \, \tau_\nu{}^B \, \epsilon_{AB}\,.
\ee
Moreover, for the purpose of this paper, we only care about $q = 0, 2, 4$\,.
After taking the limit $\omega\to\infty$, the fields $\tau_{\mu\nu}$\,, $E_{\mu\nu}$\,, $B_{\mu\nu}$\,, $\Phi$\,, and $C^{(q)}$ are interpreted as the background fields of nonrelativistic string theory that we have reviewed in Section~\ref{sec:reviewncs}.
Note that all hatted symbols represent quantities in relativistic string theory. The nonrelativistic string sigma model \eqref{eq:genaction} and nonrelativistic D-brane action \eqref{eq:dpaction} then arise from taking the $\omega \rightarrow \infty$ limit of their associated relativistic theories \cite{Andringa:2012uz,  Bergshoeff:2019pij,  Gomis:2020fui, Bidussi:2021ujm, Ebert:2021mfu}.
We emphasize that these resulting extended objects in nonrelativistic string theory are coupled to string Newton-Cartan geometry with a codimension-two foliation. This is distinct from the general $p$-brane limits considered in the literature, where a codimension-($p+1$) foliation is induced in the target space \cite{Gomis:2000bd, Gomis:2004pw}.

We now apply the nonrelativistic string limit to the manifestly \SLZ-invariant actions for relativistic instantons, strings, and D3-branes in Eq.~\eqref{eq:relbranes}.
The $\omega \rightarrow \infty$ limit of the relativistic D-instanton action \eqref{eq:relinst} proceeds in a rather straightforward way, which leads to the nonrelativistic instanton action \eqref{eq:sld-1}. In the following, we discuss how the nonrelativistic $(p\,,q)$-string and D3-brane actions arise from the nonrelativistic string limit.

Expanding the relativistic $(p\,,q)$-string action \eqref{eq:relpqstring} with respect to a large $\omega$\,, we find
\begin{align} \label{eq:divs}
    \hat S_\text{string} = -\omega^2\,T_1\,\int \ell\,\Big[|p-q\,C^{(0)}| - (p-q\,C^{(0)})\Big] + O(\omega^0)\,,
\end{align}
which implies that the $\omega \rightarrow \infty$ limit is well-defined only if the condition $p-q\,\Cz > 0$ is satisfied. Note that the leading $O(\omega^2)$ term in Eq.~\eqref{eq:divs} receives contributions from both the kinetic and the Wess-Zumino part of the relativistic action \eqref{eq:relpqstring}.
As a result,
\begin{align}      \lim\limits_{\omega\to\infty}\hat{S}^{\phantom{\dagger}}_\text{string} = S^+_\text{string}\,,\hskip 2.356truecm\mathrm{if} \,\,\, 
    p - q \, \Cz>0\,,
\end{align}
where $S^+_\text{string}$ is given in Eq.~\eqref{eq:spq}. This shows that the $\omega\to\infty$ limit effectively decouples half of the relativistic states --- namely, those with $p-q\,\Cz<0$. Moreover, taking the nonrelativistic string limit of the relativistic $(p,\,q)$-string tension  \eqref{eq:htpq}, where $\hat g_s = \omega \, g_s$ due to the rescaling of the dilaton, leads to the nonrelativistic $(p,\,q)$-string tension \eqref{eq:nrstension} for the effective action  \eqref{eq:etpqnr}.

\begin{figure}[t!]
\centering
    \begin{tikzcd}[row sep=large]
        & \text{\scalebox{1.2}{\color{orange}\,$\hat{\!\textbf{\emph{S}}}_\text{string}$}}
        \arrow[ddr, thick, dashed,
            "\text{\,\color{blue}$p-q\,C^{(0)}<0$}"]
        \arrow[ddl, thick, dashed,
            "\text{\color{blue}$p-q\,C^{(0)}>0$}"'] &\\
        & \text{\,$\omega \rightarrow \infty$} & &\\
        \text{\scalebox{1.2}{\color{orange}${\textbf{\,\emph{S}}}^\textbf{+}_\text{string}$}}
        \arrow[loop left, thick,
            "\text{\color{blue}$c\,C^{(0)}+d>0$}\,\,"]
        \arrow[rr, thick, leftrightarrow,
            "\text{\color{blue}$c\,C^{(0)}+d<0$}"'] && \text{\scalebox{1.2}{\color{orange}${\textbf{\,\emph{S}}}^\textbf{--}_\text{string}$}}
        \arrow[loop right, thick,
            "\,\,\text{\color{blue}$c\,C^{(0)}+d>0$}"]
    \end{tikzcd}
\caption{This figure illustrates the limiting procedure for nonrelativistic $(p\,,q)$-strings. The positive branch arises from the $\omega \rightarrow \infty$ limit with the reparametrization \eqref{eq:TSrescal} of the background fields in relativistic string theory, which leads to the nonrelativistic $(p\,,q)$-string action $S^+_\text{string}$\,, satisfying the condition $p -q \, C^{(0)} > 0$\,. The negative branch arises from the $\omega \rightarrow \infty$ limit with the reparametrization \eqref{eq:mirrorrepar}, which leads to the nonrelativistic $(p\,,q)$-string action $S^-_\text{string}$\,, satisfying the condition $p -q \, C^{(0)} < 0$\,. While the \SLZ\, transformations satisfying $c \, C^{(0)} + d > 0$ map $S^+_\text{string}$ ($S^-_\text{string}$) to itself, the transformations satisfying $c \, C^{(0)} + d < 0$ map between $S^+_\text{string}$ and $S^-_\text{string}$\,.}
\label{fig:s}
\end{figure}
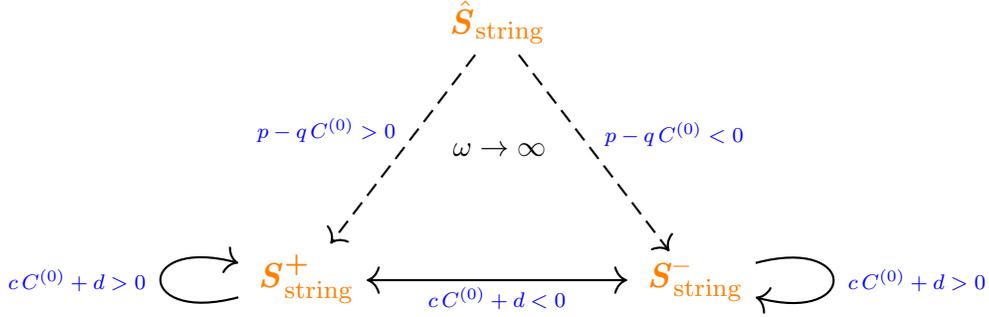

In the bulk of this paper, we have seen that the nonrelativistic string action $S^+_\text{string}$ is not invariant under the full \SLZ-duality group by themselves. Instead, under \SLZ\, transformations satisfying the condition $c \, C^{(0)} + d < 0$\,, the positive-branch action $S^+_\text{string}$ with $p - q \, C^{(0)} > 0$ is mapped to the negative-branch action $S^-_{\text{string}}$\,. It is natural to wonder whether these negative-branch actions also arise from a similar nonrelativistic string limit. This can be achieved by slightly changing the reparametrization by inverting the signs of the two-forms. Effectively, one can define the negative-branch limit by replacing the rules for the two forms in Eq.~\eqref{eq:TSrescal} by
\begin{subequations} \label{eq:mirrorrepar}
\begin{align}
    \hat B & = \omega^2 \, \ell - B\,,
        &
    \hat C^{(2)} & = -\omega^2 \, C^{(0)} \, \ell - C^{(2)}\,. \label{eq:bc2}
\end{align}
\end{subequations}
Using this parametrization, we find in the $\omega \rightarrow \infty$ limit that
\begin{align}
    \Lim{\omega\to\infty}\hat{S}^{\phantom{\dagger}}_\text{string} = S^-_\text{string}\,, 
        \hskip 2.356truecm%
    \text{if} \,\,\, p - q \, \Cz<0\,,
\end{align}
where $S^-_\text{string}$ is given in Eq.~\eqref{eq:spqm}.

We observe that the branching of the $(p\,,q)$-space naturally follows from the occurrence of a divergence in the nonrelativistic limit. 
Physically, this can be understood as follows: For a given $\hat C^{(0)}$, certain states in the $(p\,,q)$-space become infinitely heavy as $\omega\to\infty$. The $\omega \rightarrow \infty$ limit associated with the background field reparametrization \eqref{eq:TSrescal}
decouples all the states with $p-q\,\Cz<0$\,. The $\omega \rightarrow \infty$ limit associated with
the reparametrization \eqref{eq:mirrorrepar}, on the other hand, decouples all states with $p-q\,\Cz>0$. The surviving states are described by $S^+_\text{string}$ and $S^-_\text{string}$\,, respectively. This limiting procedure is exemplified in Figure \ref{fig:s}.

We note that there is a complementary but equivalent procedure where instead of taking two inequivalent limits of one action one takes one limit of two inequivalent actions. These actions differ from each other by the sign of the Wess-Zumino term.

It is also possible to reproduce the topological action \eqref{eq:tls} by starting with the relativistic $(p\,,q)$-string action \eqref{eq:relpqstring} satisfying the condition $p - q \, C^{(0)} = 0$\,.
Using the parametrizations in Eq.~\eqref{eq:TSrescal}, together with the additional ansatz,
\be
    T_1 = \frac{T_\text{eff}}{\omega}\,,
\ee
we find that the action \eqref{eq:tls} arises as the $\omega \rightarrow \infty$ limit of the relativistic action.

In contrast to nonrelativistic $(p\,, q)$-strings, applying the $\omega \rightarrow \infty$ limit to the relativistic D-instanton action \eqref{eq:relinst} and D3-brane action \eqref{eq:relD3} leads to finite results, without any requirements on $p - q \, C^{(0)}$. It then follows that
\begin{align}
    \lim_{\omega\to\infty} \hat{S}_\text{D(-1)} = S_\text{D(-1)}\,,
        \qquad%
    \lim_{\omega\to\infty}\hat{S}^{\phantom{\dagger}}_{\text{D3}} = S_{\text{D3}}\,,
\end{align}
where the nonrelativistic D-instanton action $S_\text{D(-1)}$ is given in Eq.~\eqref{eq:sld-1} and D3-brane action $S_{\text{D3}}$ is given in Eq.~\eqref{eq:sd3bb}.

Similar to the non-relativistic limit of the relativistic action we have discussed above, the non-relativistic \SLZ\  transformations can be obtained by a large $\omega$ expansion of the relativistic fields in Eq.~\eqref{eq:lgts}. The  relativistic \SLZ\ transformations in Eq.~\eqref{sltran} and the transformation of $\hat C^{(4)}$ can be expressed as
\begin{align} \label{relStr}
    \hat G_{\alpha\beta}' &= \ls (c \, \hat{C}^{(0)} + d)^2 + c^2 \, e^{-2\hat{\Phi}} \rs^{\frac{1}{2}} \, \hat{G}_{\alpha\beta} \,,
        &
    \hat B' &=  d \, \hat B  - c \, \hat C^{(2)}\,, \notag \\[10pt]
    e^{\hat \Phi '} &= \ls (c \, \hat{C}^{(0)} + d)^2 + c^2 \, e^{-2\hat{\Phi}} \rs e^{\hat \Phi}\,,
        &
    \hat C^{(2)'}  &=  a \, \hat C^{(2)} - b \, \hat B\,, \\[4pt]
    \hat C^{(0)'} &= \frac{(a \, \hat{C}^{(0)} + d) \, \bigl( c \, \hat{C}^{(0)} + d \bigr) + a \, c \, e^{-2\hat{\Phi}}}{\bigl( c \, \hat{C}^{(0)} + d \bigr)^2 + c^2 \, e^{-2\hat{\Phi}}}\,,
        &
    \hat C'{}^{(4)} & = \hat C^{(4)} + \frac{1}{2} \,  \hat \Sigma^\intercal
    \begin{pmatrix}
        b \, d &\,\,\, - b \, c \\[2pt]
        - b \, c &\,\,\, a \, c
    \end{pmatrix}
    \wedge \hat \Sigma\,. \notag
\end{align}
Substituting the expansion \eqref{eq:TSrescal} into both sides of the transformations in Eq.~\eqref{relStr}, we get the non-relativistic \SLZ \ transformation in Eqs.~\eqref{nrsltrans} and \eqref{nrsl3} for $c \, C^{(0)} + d > 0$\,, which is a required condition for the resulting branched \SLZ-transformations to be finite.

\newpage

\bibliographystyle{JHEP}
\bibliography{bsld}

\providecommand{\href}[2]{#2}\begingroup\raggedright\begin{thebibliography}{10}

\bibitem{Schwarz:1983wa}
J.~H. Schwarz and P.~C. West, \emph{{Symmetries and transformations of chiral
  N=2 D=10 supergravity}},
  \href{https://doi.org/10.1016/0370-2693(83)90168-5}{\emph{Phys. Lett. B}
  {\bfseries 126} (1983) 301--304}.

\bibitem{Hull:1994ys}
C.~M. Hull and P.~K. Townsend, \emph{{Unity of superstring dualities}},
  \href{https://doi.org/10.1016/0550-3213(94)00559-W}{\emph{Nucl. Phys. B}
  {\bfseries 438} (1995) 109--137},
  [\href{https://arxiv.org/abs/hep-th/9410167}{{\ttfamily hep-th/9410167}}].

\bibitem{Schwarz:1995dk}
J.~H. Schwarz, \emph{{An SL(2,Z) multiplet of type IIB superstrings}},
  \href{https://doi.org/10.1016/0370-2693(95)01405-5}{\emph{Phys. Lett. B}
  {\bfseries 360} (1995) 13--18},
  [\href{https://arxiv.org/abs/hep-th/9508143}{{\ttfamily hep-th/9508143}}].

\bibitem{Becker:2006dvp}
K.~Becker, M.~Becker and J.~H. Schwarz, \emph{{String theory and M-theory: A
  modern introduction}}.
\newblock Cambridge University Press, 12, 2006,
  \href{https://doi.org/10.1017/CBO9780511816086}{10.1017/CBO9780511816086}.

\bibitem{cardy1982duality}
J.~L. Cardy, \emph{Duality and the $\theta$ parameter in abelian lattice
  models}, {\emph{Nuclear Physics B} {\bfseries 205} (1982) 17--26}.

\bibitem{cardy1982phase}
J.~L. Cardy and E.~Rabinovici, \emph{Phase structure of zp models in the
  presence of a $\theta$ parameter}, {\emph{Nuclear Physics B} {\bfseries 205}
  (1982) 1--16}.

\bibitem{shapere1989self}
A.~Shapere and F.~Wilczek, \emph{Self-dual models with theta terms},
  {\emph{Nuclear Physics B} {\bfseries 320} (1989) 669--695}.

\bibitem{Townsend:1997kr}
P.~Townsend, \emph{{Membrane tension and manifest IIB S duality}},
  \href{https://doi.org/10.1016/S0370-2693(97)00862-9}{\emph{Phys. Lett. B}
  {\bfseries 409} (1997) 131--135},
  [\href{https://arxiv.org/abs/hep-th/9705160}{{\ttfamily hep-th/9705160}}].

\bibitem{Cederwall:1997ts}
M.~Cederwall and P.~K. Townsend, \emph{{The manifestly Sl(2,Z) covariant
  superstring}},
  \href{https://doi.org/10.1088/1126-6708/1997/09/003}{\emph{JHEP} {\bfseries
  09} (1997) 003}, [\href{https://arxiv.org/abs/hep-th/9709002}{{\ttfamily
  hep-th/9709002}}].

\bibitem{Tseytlin:1996it}
A.~A. Tseytlin, \emph{{Self-duality of Born-Infeld action and Dirichlet
  three-brane of type IIB superstring theory}},
  \href{https://doi.org/10.1016/0550-3213(96)00173-3}{\emph{Nucl. Phys. B}
  {\bfseries 469} (1996) 51--67},
  [\href{https://arxiv.org/abs/hep-th/9602064}{{\ttfamily hep-th/9602064}}].

\bibitem{Bergshoeff:2006gs}
E.~A. Bergshoeff, M.~de~Roo, S.~F. Kerstan, T.~Ort\'{i}n and F.~Riccioni,
  \emph{{SL(2,R)-invariant IIB brane actions}},
  \href{https://doi.org/10.1088/1126-6708/2007/02/007}{\emph{JHEP} {\bfseries
  02} (2007) 007}, [\href{https://arxiv.org/abs/hep-th/0611036}{{\ttfamily
  hep-th/0611036}}].

\bibitem{Klebanov:2000pp}
I.~R. Klebanov and J.~M. Maldacena, \emph{{(1+1)-dimensional NCOS and its U(N)
  gauge theory dual}},
  \href{https://doi.org/10.1142/S0217751X01004001}{\emph{Adv. Theor. Math.
  Phys.} {\bfseries 4} (2000) 283--302},
  [\href{https://arxiv.org/abs/hep-th/0006085}{{\ttfamily hep-th/0006085}}].

\bibitem{Gomis:2000bd}
J.~Gomis and H.~Ooguri, \emph{{Nonrelativistic closed string theory}},
  \href{https://doi.org/10.1063/1.1372697}{\emph{J. Math. Phys.} {\bfseries 42}
  (2001) 3127--3151}, [\href{https://arxiv.org/abs/hep-th/0009181}{{\ttfamily
  hep-th/0009181}}].

\bibitem{Danielsson:2000gi}
U.~H. Danielsson, A.~Guijosa and M.~Kruczenski, \emph{{IIA/B, wound and
  wrapped}}, \href{https://doi.org/10.1088/1126-6708/2000/10/020}{\emph{JHEP}
  {\bfseries 10} (2000) 020},
  [\href{https://arxiv.org/abs/hep-th/0009182}{{\ttfamily hep-th/0009182}}].

\bibitem{Ebert:2021mfu}
S.~Ebert, H.-Y. Sun and Z.~Yan, \emph{{Dual D-brane actions in nonrelativistic
  string theory}}, \href{https://doi.org/10.1007/JHEP04(2022)161}{\emph{JHEP}
  {\bfseries 04} (2022) 161},
  [\href{https://arxiv.org/abs/arXiv:2112.09316}{{\ttfamily
  arXiv:2112.09316}}].

\bibitem{Bergshoeff:2018yvt}
E.~Bergshoeff, J.~Gomis and Z.~Yan, \emph{{Nonrelativistic string theory and
  T-duality}}, \href{https://doi.org/10.1007/JHEP11(2018)133}{\emph{JHEP}
  {\bfseries 11} (2018) 133},
  [\href{https://arxiv.org/abs/arXiv:1806.06071}{{\ttfamily
  arXiv:1806.06071}}].

\bibitem{Banks:1996vh}
T.~Banks, W.~Fischler, S.~H. Shenker and L.~Susskind, \emph{{M theory as a
  matrix model: A conjecture}},
  \href{https://doi.org/10.1103/PhysRevD.55.5112}{\emph{Phys. Rev. D}
  {\bfseries 55} (1997) 5112--5128},
  [\href{https://arxiv.org/abs/hep-th/9610043}{{\ttfamily hep-th/9610043}}].

\bibitem{Susskind:1997cw}
L.~Susskind, \emph{{Another conjecture about M(atrix) theory}},
  \href{https://arxiv.org/abs/hep-th/9704080}{{\ttfamily hep-th/9704080}}.

\bibitem{Seiberg:1997ad}
N.~Seiberg, \emph{{Why is the matrix model correct?}},
  \href{https://doi.org/10.1103/PhysRevLett.79.3577}{\emph{Phys. Rev. Lett.}
  {\bfseries 79} (1997) 3577--3580},
  [\href{https://arxiv.org/abs/hep-th/9710009}{{\ttfamily hep-th/9710009}}].

\bibitem{Sen:1997we}
A.~Sen, \emph{{D0-branes on T${}^{ n}$ and matrix theory}},
  \href{https://doi.org/10.4310/ATMP.1998.v2.n1.a2}{\emph{Adv. Theor. Math.
  Phys.} {\bfseries 2} (1998) 51--59},
  [\href{https://arxiv.org/abs/hep-th/9709220}{{\ttfamily hep-th/9709220}}].

\bibitem{Andringa:2012uz}
R.~Andringa, E.~Bergshoeff, J.~Gomis and M.~de~Roo, \emph{{`Stringy'
  Newton-Cartan gravity}},
  \href{https://doi.org/10.1088/0264-9381/29/23/235020}{\emph{Class. Quant.
  Grav.} {\bfseries 29} (2012) 235020},
  [\href{https://arxiv.org/abs/arXiv:1206.5176}{{\ttfamily arXiv:1206.5176}}].

\bibitem{Harmark:2017rpg}
T.~Harmark, J.~Hartong and N.~A. Obers, \emph{{Nonrelativistic strings and
  limits of the AdS/CFT correspondence}},
  \href{https://doi.org/10.1103/PhysRevD.96.086019}{\emph{Phys. Rev. D}
  {\bfseries 96} (2017) 086019},
  [\href{https://arxiv.org/abs/arXiv:1705.03535}{{\ttfamily
  arXiv:1705.03535}}].

\bibitem{Harmark:2018cdl}
T.~Harmark, J.~Hartong, L.~Menculini, N.~A. Obers and Z.~Yan, \emph{{Strings
  with non-relativistic conformal symmetry and limits of the AdS/CFT
  correspondence}}, \href{https://doi.org/10.1007/JHEP11(2018)190}{\emph{JHEP}
  {\bfseries 11} (2018) 190},
  [\href{https://arxiv.org/abs/arXiv:1810.05560}{{\ttfamily
  arXiv:1810.05560}}].

\bibitem{Harmark:2019upf}
T.~Harmark, J.~Hartong, L.~Menculini, N.~A. Obers and G.~Oling, \emph{{Relating
  non-relativistic string theories}},
  \href{https://doi.org/10.1007/JHEP11(2019)071}{\emph{JHEP} {\bfseries 11}
  (2019) 071}, [\href{https://arxiv.org/abs/arXiv:1907.01663}{{\ttfamily
  arXiv:1907.01663}}].

\bibitem{Bergshoeff:2019pij}
E.~A. Bergshoeff, J.~Gomis, J.~Rosseel, C.~\c{S}im\c{s}ek and Z.~Yan,
  \emph{{String theory and string Newton-Cartan geometry}},
  \href{https://doi.org/10.1088/1751-8121/ab56e9}{\emph{J. Phys. A} {\bfseries
  53} (2020) 014001}, [\href{https://arxiv.org/abs/arXiv:1907.10668}{{\ttfamily
  arXiv:1907.10668}}].

\bibitem{Kluson:2019ifd}
J.~Kluso\v{n}, \emph{{$(m,n)$-string and D1-brane in stringy Newton-Cartan
  background}}, \href{https://doi.org/10.1007/JHEP04(2019)163}{\emph{JHEP}
  {\bfseries 04} (2019) 163},
  [\href{https://arxiv.org/abs/arXiv:1901.11292}{{\ttfamily
  arXiv:1901.11292}}].

\bibitem{Kluson:2019avy}
J.~Kluso\v{n}, \emph{{Non-relativistic D-brane from T-duality along null
  direction}}, \href{https://doi.org/10.1007/JHEP10(2019)153}{\emph{JHEP}
  {\bfseries 10} (2019) 153},
  [\href{https://arxiv.org/abs/arXiv:1907.05662}{{\ttfamily
  arXiv:1907.05662}}].

\bibitem{Gallegos:2020egk}
A.~D. Gallegos, U.~G\"ursoy, S.~Verma and N.~Zinnato, \emph{{Non-Riemannian
  gravity actions from Double Field Theory}},
  \href{https://doi.org/10.1007/JHEP06(2021)173}{\emph{JHEP} {\bfseries 06}
  (2021) 173}, [\href{https://arxiv.org/abs/arXiv:2012.07765}{{\ttfamily
  arXiv:2012.07765}}].

\bibitem{Gomis:2020fui}
J.~Gomis, Z.~Yan and M.~Yu, \emph{{Nonrelativistic open string and Yang-Mills
  theory}}, \href{https://doi.org/10.1007/JHEP03(2021)269}{\emph{JHEP}
  {\bfseries 03} (2021) 269},
  [\href{https://arxiv.org/abs/arXiv:2007.01886}{{\ttfamily
  arXiv:2007.01886}}].

\bibitem{Bergshoeff:2021bmc}
E.~A. Bergshoeff, J.~Lahnsteiner, L.~Romano, J.~Rosseel and C.~\c{S}im\c{s}ek,
  \emph{{A non-relativistic limit of NS-NS gravity}},
  \href{https://doi.org/10.1007/JHEP06(2021)021}{\emph{JHEP} {\bfseries 06}
  (2021) 021}, [\href{https://arxiv.org/abs/arXiv:2102.06974}{{\ttfamily
  arXiv:2102.06974}}].

\bibitem{Bidussi:2021ujm}
L.~Bidussi, T.~Harmark, J.~Hartong, N.~A. Obers and G.~Oling, \emph{{Torsional
  string Newton-Cartan geometry for non-relativistic strings}},
  \href{https://arxiv.org/abs/arXiv:2107.00642}{{\ttfamily arXiv:2107.00642}}.

\bibitem{Bergshoeff:2021tfn}
E.~A. Bergshoeff, J.~Lahnsteiner, L.~Romano, J.~Rosseel and C.~\c{S}im\c{s}ek,
  \emph{{Non-relativistic ten-dimensional minimal supergravity}},
  \href{https://arxiv.org/abs/arXiv:2107.14636}{{\ttfamily arXiv:2107.14636}}.

\bibitem{Bergshoeff:2022pzk}
E.~Bergshoeff, J.~Lahnsteiner, L.~Romano and J.~Rosseel, \emph{{The
  supersymmetric Neveu-Schwarz branes of non-relativistic string theory}},
  \href{https://doi.org/10.1007/JHEP08(2022)218}{\emph{JHEP} {\bfseries 08}
  (2022) 218}, [\href{https://arxiv.org/abs/arXiv:2204.04089}{{\ttfamily
  arXiv:2204.04089}}].

\bibitem{Oling:2022fft}
G.~Oling and Z.~Yan, \emph{{Aspects of nonrelativistic strings}},
  \href{https://doi.org/10.3389/fphy.2022.832271}{\emph{Front. in Phys.}
  {\bfseries 10} (2022) 832271},
  [\href{https://arxiv.org/abs/arXiv:2202.12698}{{\ttfamily
  arXiv:2202.12698}}].

\bibitem{Gomis:2019zyu}
J.~Gomis, J.~Oh and Z.~Yan, \emph{{Nonrelativistic string theory in background
  fields}}, \href{https://doi.org/10.1007/JHEP10(2019)101}{\emph{JHEP}
  {\bfseries 10} (2019) 101},
  [\href{https://arxiv.org/abs/arXiv:1905.07315}{{\ttfamily
  arXiv:1905.07315}}].

\bibitem{Gallegos:2019icg}
A.~D. Gallegos, U.~G\"ursoy and N.~Zinnato, \emph{{Torsional Newton Cartan
  gravity from non-relativistic strings}},
  \href{https://doi.org/10.1007/JHEP09(2020)172}{\emph{JHEP} {\bfseries 09}
  (2020) 172}, [\href{https://arxiv.org/abs/arXiv:1906.01607}{{\ttfamily
  arXiv:1906.01607}}].

\bibitem{Yan:2019xsf}
Z.~Yan and M.~Yu, \emph{{Background field method for nonlinear sigma models in
  nonrelativistic string theory}},
  \href{https://doi.org/10.1007/JHEP03(2020)181}{\emph{JHEP} {\bfseries 03}
  (2020) 181}, [\href{https://arxiv.org/abs/arXiv:1912.03181}{{\ttfamily
  arXiv:1912.03181}}].

\bibitem{Yan:2021lbe}
Z.~Yan, \emph{{Torsional deformation of nonrelativistic string theory}},
  \href{https://arxiv.org/abs/arXiv:2106.10021}{{\ttfamily arXiv:2106.10021}}.

\bibitem{Gopakumar:2000na}
R.~Gopakumar, J.~M. Maldacena, S.~Minwalla and A.~Strominger, \emph{{S duality
  and noncommutative gauge theory}},
  \href{https://doi.org/10.1088/1126-6708/2000/06/036}{\emph{JHEP} {\bfseries
  06} (2000) 036}, [\href{https://arxiv.org/abs/hep-th/0005048}{{\ttfamily
  hep-th/0005048}}].

\bibitem{Gomis:2020izd}
J.~Gomis, Z.~Yan and M.~Yu, \emph{{T-duality in nonrelativistic open string
  theory}}, \href{https://doi.org/10.1007/JHEP02(2021)087}{\emph{JHEP}
  {\bfseries 02} (2021) 087},
  [\href{https://arxiv.org/abs/arXiv:2008.05493}{{\ttfamily
  arXiv:2008.05493}}].

\bibitem{Russo:2000zb}
J.~G. Russo and M.~M. Sheikh-Jabbari, \emph{{On noncommutative open string
  theories}}, \href{https://doi.org/10.1088/1126-6708/2000/07/052}{\emph{JHEP}
  {\bfseries 07} (2000) 052},
  [\href{https://arxiv.org/abs/hep-th/0006202}{{\ttfamily hep-th/0006202}}].

\bibitem{Cai:2000yk}
R.-G. Cai and N.~Ohta, \emph{{(F1, D1, D3) bound state, its scaling limits and
  SL(2,Z) duality}}, \href{https://doi.org/10.1143/PTP.104.1073}{\emph{Prog.
  Theor. Phys.} {\bfseries 104} (2000) 1073--1087},
  [\href{https://arxiv.org/abs/hep-th/0007106}{{\ttfamily hep-th/0007106}}].

\bibitem{Lu:2000vv}
J.~X. Lu, S.~Roy and H.~Singh, \emph{{SL(2, Z) duality and four-dimensional
  noncommutative theories}},
  \href{https://doi.org/10.1016/S0550-3213(00)00671-4}{\emph{Nucl. Phys. B}
  {\bfseries 595} (2001) 298--318},
  [\href{https://arxiv.org/abs/hep-th/0007168}{{\ttfamily hep-th/0007168}}].

\bibitem{Gran:2001tk}
U.~Gran and M.~Nielsen, \emph{{Noncommutative open (p,\,q) string theories}},
  \href{https://doi.org/10.1088/1126-6708/2001/11/022}{\emph{JHEP} {\bfseries
  11} (2001) 022}, [\href{https://arxiv.org/abs/hep-th/0104168}{{\ttfamily
  hep-th/0104168}}].

\bibitem{Gomis:2005pg}
J.~Gomis, J.~Gomis and K.~Kamimura, \emph{{Non-relativistic superstrings: A new
  soluble sector of AdS$_5\times S^5$}},
  \href{https://doi.org/10.1088/1126-6708/2005/12/024}{\emph{JHEP} {\bfseries
  12} (2005) 024}, [\href{https://arxiv.org/abs/hep-th/0507036}{{\ttfamily
  hep-th/0507036}}].

\bibitem{Danielsson:2000mu}
U.~H. Danielsson, A.~Guijosa and M.~Kruczenski, \emph{{Newtonian gravitons and
  D-brane collective coordinates in wound string theory}},
  \href{https://doi.org/10.1088/1126-6708/2001/03/041}{\emph{JHEP} {\bfseries
  03} (2001) 041}, [\href{https://arxiv.org/abs/hep-th/0012183}{{\ttfamily
  hep-th/0012183}}].

\bibitem{Yan:2021hte}
Z.~Yan and M.~Yu, \emph{{KLT factorization of nonrelativistic string
  amplitudes}}, \href{https://doi.org/10.1007/JHEP04(2022)068}{\emph{JHEP}
  {\bfseries 04} (2022) 068},
  [\href{https://arxiv.org/abs/arXiv:2112.00025}{{\ttfamily
  arXiv:2112.00025}}].

\bibitem{Townsend:1996xj}
P.~K. Townsend, \emph{{Four lectures on M theory}},  in \emph{{ICTP Summer
  School in High-energy Physics and Cosmology}}, pp.~385--438, 12, 1996,
  \href{https://arxiv.org/abs/hep-th/9612121}{{\ttfamily hep-th/9612121}}.

\bibitem{Aganagic:1997zk}
M.~Aganagic, J.~Park, C.~Popescu and J.~H. Schwarz, \emph{{Dual D-brane
  actions}}, \href{https://doi.org/10.1016/S0550-3213(97)00257-5}{\emph{Nucl.
  Phys. B} {\bfseries 496} (1997) 215--230},
  [\href{https://arxiv.org/abs/hep-th/9702133}{{\ttfamily hep-th/9702133}}].

\bibitem{Dabholkar:1989jt}
A.~Dabholkar and J.~A. Harvey, \emph{{Nonrenormalization of the superstring
  tension}}, \href{https://doi.org/10.1103/PhysRevLett.63.478}{\emph{Phys. Rev.
  Lett.} {\bfseries 63} (1989) 478}.

\bibitem{Seiberg:1999vs}
N.~Seiberg and E.~Witten, \emph{{String theory and noncommutative geometry}},
  \href{https://doi.org/10.1088/1126-6708/1999/09/032}{\emph{JHEP} {\bfseries
  09} (1999) 032}, [\href{https://arxiv.org/abs/hep-th/9908142}{{\ttfamily
  hep-th/9908142}}].

\bibitem{Gubser:2000mf}
S.~S. Gubser, S.~Gukov, I.~R. Klebanov, M.~Rangamani and E.~Witten, \emph{{The
  Hagedorn transition in noncommutative open string theory}},
  \href{https://doi.org/10.1063/1.1372176}{\emph{J. Math. Phys.} {\bfseries 42}
  (2001) 2749--2764}, [\href{https://arxiv.org/abs/hep-th/0009140}{{\ttfamily
  hep-th/0009140}}].

\bibitem{Meessen:1998qm}
P.~Meessen and T.~Ort\'{i}n, \emph{{An Sl(2,Z) multiplet of nine-dimensional
  type II supergravity theories}},
  \href{https://doi.org/10.1016/S0550-3213(98)00780-9}{\emph{Nucl. Phys. B}
  {\bfseries 541} (1999) 195--245},
  [\href{https://arxiv.org/abs/hep-th/9806120}{{\ttfamily hep-th/9806120}}].

\bibitem{Gomis:2004pw}
J.~Gomis, K.~Kamimura and P.~K. Townsend, \emph{{Non-relativistic
  superbranes}},
  \href{https://doi.org/10.1088/1126-6708/2004/11/051}{\emph{JHEP} {\bfseries
  11} (2004) 051}, [\href{https://arxiv.org/abs/hep-th/0409219}{{\ttfamily
  hep-th/0409219}}].

\end{thebibliography}\endgroup

\end{document}